\documentclass[fleqn,usenatbib]{mnras}

\usepackage{newtxtext,newtxmath}
\usepackage[T1]{fontenc}

\DeclareRobustCommand{\VAN}[3]{#2}
\let\VANthebibliography\thebibliography
\def\thebibliography{\DeclareRobustCommand{\VAN}[3]{##3}\VANthebibliography}

\usepackage{graphicx}	% Including figure files
\usepackage{subfig}
\usepackage{amsmath}	% Advanced maths commands
\usepackage{array}
\usepackage{bm}
\usepackage{float}

\usepackage{lineno}  % For line numbers
\newcommand{\ha}{\textrm{H}\ensuremath{\alpha}}
\newcommand{\hb}{\textrm{H}\ensuremath{\beta}}

\newcommand{\oiii}{[\textrm{O}~\textsc{iii}]}
\newcommand{\oii}{[\textrm{O}~\textsc{ii}]}
\newcommand{\nii}{[\textrm{N}~\textsc{ii}]}

\newcommand{\sii}{[\textrm{S}~\textsc{ii}]}
\newcommand{\caii}{\textrm{Ca}~\textsc{ii}}
\newcommand{\nai}{\textrm{Na}~\textsc{i}}

\newcommand{\hei}{\textrm{He}~\textsc{i}}
\newcommand{\mgi}{\textrm{Mg}~\textsc{i}}
\newcommand{\mgii}{\textrm{Mg}~\textsc{ii}}
\newcommand{\ciii}{\textrm{C}~\textsc{iii] }}
\newcommand{\civ}{\textrm{C}~\textsc{iv}}
\newcommand{\oiiilam}{[\textrm{O}~\textsc{iii}]~\ensuremath{\lambda4960,\lambda5008}}

\newcommand{\oiilam}{[\textrm{O}~\textsc{ii}]~\ensuremath{\lambda3727,\lambda3730}}

\newcommand{\niidoub}{[\textrm{N}~\textsc{ii}]~\ensuremath{\lambda6550,\lambda6585}}
\newcommand{\caiihk}{\textrm{Ca}~\textsc{ii}~\textrm{K and H}~\ensuremath{\lambda3935,\lambda3970}}

\newcommand{\heilam}{\textrm{He}~\textsc{i}~\ensuremath{\lambda3889}}
\newcommand{\neiiilam}{[\textrm{Ne}~\textsc{iii}]~\ensuremath{\lambda3870}}

\newcommand{\dlines}{\textrm{Na}~\textsc{i}~\textrm{D-lines}~\ensuremath{\lambda5896,\lambda5898}}   
\newcommand{\siilam}{[\textrm{S}~\textsc{ii}]~\ensuremath{\lambda6718,\lambda6733}}
\newcommand{\halam}{\textrm{H}\ensuremath{\alpha}~\ensuremath{\lambda6565}}
\newcommand{\hblam}{\textrm{H}\ensuremath{\beta}~\ensuremath{\lambda4863}}

\title[Identifying Anomalous DESI Galaxy Spectra with a VAE]{Identifying Anomalous DESI Galaxy Spectra with a Variational Autoencoder}

\author[C. Nicolaou et al.]{
C.~Nicolaou,$^{1}$\thanks{E-mail: constantina.nicolaou.17@ucl.ac.uk}
R.P.~Nathan,$^{1}$
O.~Lahav,$^{1}$
A.~Palmese,${^2}$
A.~Saintonge,$^{1}$ 
J.~Aguilar,$^{3}$
S.~Ahlen,$^{4}$
\newauthor C.~Allende~Prieto,$^{5,}$$^{6}$
S.~Bailey,$^{3}$
S.~BenZvi,$^{7}$
D.~Bianchi,$^{8,}$$^{9}$
A.~Brodzeller,$^{3}$
D.~Brooks,$^{1}$
T.~Claybaugh,$^{3}$
\newauthor A.~de la Macorra,$^{10}$
J.~Della~Costa,$^{11,}$$^{12}$
Arjun~Dey,$^{12}$
P.~Doel,$^{1}$
J.~E.~Forero-Romero,$^{13,}$$^{14}$
\newauthor E.~Gazta\~naga,$^{15,}$$^{16,}$$^{17}$
S.~Gontcho A Gontcho,$^{3}$
G.~Gutierrez,$^{18}$
K.~Honscheid,$^{19,}$$^{20,}$$^{21}$
C.~Howlett,$^{22}$
\newauthor M.~Ishak,$^{23}$
R.~Kehoe,$^{24}$
D.~Kirkby,$^{25}$
T.~Kisner,$^{3}$
A.~Kremin,$^{3}$
A.~Lambert,$^{3}$
M.~Landriau,$^{3}$
L.~Le~Guillou,$^{26}$
\newauthor A.~Meisner,$^{12}$
R.~Miquel,$^{27,}$$^{28}$
J.~Moustakas,$^{29}$
S.~Nadathur,$^{16}$
F.~Prada,$^{30}$
I.~P\'erez-R\`afols,$^{31}$
G.~Rossi,$^{32}$
\newauthor E.~Sanchez,$^{33}$
M.~Schubnell,$^{34,}$$^{35}$
M.~Siudek,$^{6,}$$^{17}$
D.~Sprayberry,$^{12}$
G.~Tarl\'{e},$^{35}$
B.~A.~Weaver,$^{12}$
and H.~Zou,$^{36}$
\\
$^{1}$Department of Physics \& Astronomy, University College London, Gower Street, London, WC1E 6BT, UK\\
$^{2}$Department of Physics, Carnegie Mellon University, 5000 Forbes Avenue, Pittsburgh, PA 15213, USA\\
$^{3}$Lawrence Berkeley National Laboratory, 1 Cyclotron Road, Berkeley, CA 94720, USA\\
$^{4}$Physics Dept., Boston University, 590 Commonwealth Avenue, Boston, MA 02215, USA\\
$^{5}$Departamento de Astrof\'{\i}sica, Universidad de La Laguna (ULL), E-38206, La Laguna, Tenerife, Spain\\
$^{6}$Instituto de Astrof\'{\i}sica de Canarias, C/ V\'{\i}a L\'{a}ctea, s/n, E-38205 La Laguna, Tenerife, Spain\\
$^{7}$Department of Physics \& Astronomy, University of Rochester, 206 Bausch and Lomb Hall, P.O. Box 270171, Rochester, NY 14627-0171, USA\\
$^{8}$Dipartimento di Fisica ``Aldo Pontremoli'', Universit\`a degli Studi di Milano, Via Celoria 16, I-20133 Milano, Italy\\
$^{9}$INAF-Osservatorio Astronomico di Brera, Via Brera 28, 20122 Milano, Italy\\
$^{10}$Instituto de F\'{\i}sica, Universidad Nacional Aut\'{o}noma de M\'{e}xico,  Circuito de la Investigaci\'{o}n Cient\'{\i}fica, Ciudad Universitaria, Cd. de M\'{e}xico  C.~P.~04510,  M\'{e}xico\\
$^{11}$Department of Astronomy, San Diego State University, 5500 Campanile Drive, San Diego, CA 92182, USA\\
$^{12}$NSF NOIRLab, 950 N. Cherry Ave., Tucson, AZ 85719, USA\\
$^{13}$Departamento de F\'isica, Universidad de los Andes, Cra. 1 No. 18A-10, Edificio Ip, CP 111711, Bogot\'a, Colombia\\
$^{14}$Observatorio Astron\'omico, Universidad de los Andes, Cra. 1 No. 18A-10, Edificio H, CP 111711 Bogot\'a, Colombia\\
$^{15}$Institut d'Estudis Espacials de Catalunya (IEEC), c/ Esteve Terradas 1, Edifici RDIT, Campus PMT-UPC, 08860 Castelldefels, Spain\\
$^{16}$Institute of Cosmology and Gravitation, University of Portsmouth, Dennis Sciama Building, Portsmouth, PO1 3FX, UK\\
$^{17}$Institute of Space Sciences, ICE-CSIC, Campus UAB, Carrer de Can Magrans s/n, 08913 Bellaterra, Barcelona, Spain\\
$^{18}$Fermi National Accelerator Laboratory, PO Box 500, Batavia, IL 60510, USA\\
$^{19}$Center for Cosmology and AstroParticle Physics, The Ohio State University, 191 West Woodruff Avenue, Columbus, OH 43210, USA\\
$^{20}$Department of Physics, The Ohio State University, 191 West Woodruff Avenue, Columbus, OH 43210, USA\\
$^{21}$The Ohio State University, Columbus, 43210 OH, USA\\
$^{22}$School of Mathematics and Physics, University of Queensland, Brisbane, QLD 4072, Australia\\
$^{23}$Department of Physics, The University of Texas at Dallas, 800 W. Campbell Rd., Richardson, TX 75080, USA\\
$^{24}$Department of Physics, Southern Methodist University, 3215 Daniel Avenue, Dallas, TX 75275, USA\\
$^{25}$Department of Physics and Astronomy, University of California, Irvine, 92697, USA\\
$^{26}$Sorbonne Universit\'{e}, CNRS/IN2P3, Laboratoire de Physique Nucl\'{e}aire et de Hautes Energies (LPNHE), FR-75005 Paris, France\\
$^{27}$Instituci\'{o} Catalana de Recerca i Estudis Avan\c{c}ats, Passeig de Llu\'{\i}s Companys, 23, 08010 Barcelona, Spain\\
$^{28}$Institut de F\'{i}sica d'Altes Energies (IFAE), The Barcelona Institute of Science and Technology, Edifici Cn, Campus UAB, 08193, Bellaterra (Barcelona), Spain\\
$^{29}$Department of Physics and Astronomy, Siena College, 515 Loudon Road, Loudonville, NY 12211, USA\\
$^{30}$Instituto de Astrof\'{i}sica de Andaluc\'{i}a (CSIC), Glorieta de la Astronom\'{i}a, s/n, E-18008 Granada, Spain\\
$^{31}$Departament de F\'isica, EEBE, Universitat Polit\`ecnica de Catalunya, c/Eduard Maristany 10, 08930 Barcelona, Spain\\
$^{32}$Department of Physics and Astronomy, Sejong University, 209 Neungdong-ro, Gwangjin-gu, Seoul 05006, Republic of Korea\\
$^{33}$CIEMAT, Avenida Complutense 40, E-28040 Madrid, Spain\\
$^{34}$Department of Physics, University of Michigan, 450 Church Street, Ann Arbor, MI 48109, USA\\
$^{35}$University of Michigan, 500 S. State Street, Ann Arbor, MI 48109, USA\\
$^{36}$National Astronomical Observatories, Chinese Academy of Sciences, A20 Datun Rd., Chaoyang District, Beijing, 100012, P.R. China
}

\date{Accepted 2025 December 22. Received 2025 November 10; in original form 2025 May 9}

\pubyear{2026}

\begin{document}
\label{firstpage}
\pagerange{\pageref{firstpage}--\pageref{lastpage}}
\maketitle

%\linenumbers  % Adding in linenumbers for review purposes
\clearpage

\begin{abstract}
%Large galaxy surveys, play an instrumental role in EM observations of multi-messenger events and studies of their properties. DESI is expected to observe $35$ million galaxies.

\noindent The tens of millions of spectra being captured by the Dark Energy Spectroscopic Instrument (DESI) provide tremendous discovery potential. In this work we show how Machine Learning, in particular Variational Autoencoders (VAE), can detect anomalies in a sample of approximately 200,000 DESI spectra comprising galaxies, quasars and stars. We demonstrate that the VAE can compress the dimensionality of a spectrum by a factor of 100, while still retaining enough information to accurately reconstruct spectral features. We detect anomalous spectra as those with high reconstruction error and those which are isolated in the VAE latent representation. The anomalies identified fall into two categories: spectra with artefacts and spectra with unique physical features. Awareness of the former could improve the DESI spectroscopic pipeline; whilst the latter could help us discover new and unusual objects. 
To further curate the list of outliers identified, we use the Astronomaly package which employs Active Learning to provide personalised outlier recommendations for visual inspection. In this work we also explore the VAE latent space, finding that different object classes and subclasses are separated despite being unlabelled. 
We inject controlled synthetic anomalies and analyse their locations in the latent space to illustrate how the VAE responds to atypical spectral features;
and we demonstrate the interpretability of this latent space by identifying tracks within it that correspond to various spectral characteristics. 
%For example, we find tracks that correspond to increasing star formation and increase in broad emission lines along the Balmer series. 
In upcoming work we hope to apply the methods presented here to search for both systematics and astrophysically interesting objects in much larger datasets of DESI spectra.

% This is a simple template for authors to write new MNRAS papers.
% The abstract should briefly describe the aims, methods, and main results of the paper.
% It should be a single paragraph not more than 250 words (200 words for Letters).
% No references should appear in the abstract.
\end{abstract}


%%%%%%%%%%%%%%%%%%%%%%%%%%%%%%%%%%%%%%%%%%%%%%%%%%

%%%%%%%%%%%%%%%%% BODY OF PAPER %%%%%%%%%%%%%%%%%%

\section{Introduction}

Astronomical spectra are highly informative,
%Apart from accurate redshifts, 
as they provide a way to infer the composition, temperature and physical processes that occur in the observed object. Galaxy spectra are crucial for understanding the properties and evolution of galaxies. 
Spectroscopic surveys are currently gathering such data at an unprecedented rate providing astronomers with a wealth of information. The Dark Energy Spectroscopic Instrument \citep[DESI,][]{DESI2016a.Science} has already gathered $\sim 40$ million spectra (including repeat observations) surpassing the number of spectra previously available, aiming at an eventual total of around 35 million unique spectra. The increased volume of incoming data have accelerated the shift from traditional data analysis and visualisation tools \citep[e.g.][]{Lahav_1995} to big-data techniques \citep[e.g.][]{Baron2019ML, HuertasLanusseMLReview2022}. 

Hidden within these huge new datasets are likely to be large numbers of anomalies. An anomaly is defined as an object or observation that is rare and differs significantly from the majority of the other data accompanying it, suggesting that it has been generated by a different underlying mechanism \citep{Hawkins1980AD}. Anomalies can be rarely occurring events, extreme objects, errors due to instrument malfunction, or objects that are completely new and have the potential to unveil new physics. The last of these, also sometimes referred to as "unknown unknowns", are arguably the most interesting but also most challenging to find in large datasets. Historically, astronomers relied on visual inspection of data to make such scientific discoveries, but as volumes of data increase, visual inspection alone becomes infeasible. Astronomers must therefore turn to automated tools that can provide a smaller, curated subset of the data containing those observations most likely to be outliers.

In particular, machine learning has become increasingly popular as a powerful tool for automating the detection of atypical or anomalous observations. 
Two types of anomaly detection that are important to differentiate are outlier detection and novelty detection. Outlier detection refers to the case where the training data contains both typical data (inliers), which constitute the majority of the dataset, and anomalous data (outliers). The goal is to identify the outliers present in the dataset. In novelty detection, the training dataset contains only typical data, and the goal is to identify whether a new observation is an outlier. In this context an outlier is also called a novelty. Novelty detection is semi-supervised whereas outlier detection is unsupervised \citep{Ruff2020AD}.

Three broad categories of anomaly detection methods are deviation based, proximity based and statistical. Deviation-based anomaly detection uses the reconstruction error from a model trained on the data as the anomaly score. The model learns a lower-dimensional \textit{latent} representation of the data and then attempts to use this to reconstruct the orginal. Anomalies will tend to have a high reconstruction error as they have features that are atypical and which the model is not able to reproduce. Anomaly-detection methods based on proximity assume that anomalous data are isolated in the latent representation from the majority of the data. Approaches to define how isolated a sample is include clustering, distance-based measurements and density estimation. Statistical anomaly detection makes use of parametric or non-parametric models to define a probability distribution assumed to describe the data. A sample is anomalous if it has a low probability of being generated by the model \citep{AnCho2015VaeAD, Reis2019Unsuprv}. In this work, we make use of both deviation-based and proximity-based methods.

Deviation-based methods for spectra rely on the intuition common to many fields of study, that though the data are typically captured in high dimension, they can be represented with good fidelity using a much lower-dimensional manifold (the "manifold hypothesis"). For example, DESI spectra have $\sim 7800$ bins (spectral channels) and therefore effectively $7800$ dimensions. Due to the astrophysical processes underlying the spectra, spectral lines and features like the continuum are highly correlated. Deviation-based methods consequently tend to rely on dimensionality-reduction techniques. Because of their simplified representation of the data being studied, such techniques often have the added benefit of providing an unsupervised encoding of significant physical characteristics.

Principal Component Analysis (PCA) has been one of the most widely and successfully used algorithms for dimensionality reduction and for the unsupervised extraction of useful features from spectroscopic data. 
\cite{Boroson_2010} used the PCA reconstruction error to detect anomalies in a sample of Sloan Digital Sky Survey (SDSS) quasar spectra. 
\cite{Folkes1999} and \cite{Madgwick2003} used PCA to effectively compress galaxy spectra to a few components, finding correlations between the PCA components and physical attributes such as star formation rate and morphological type. Spectral classification derived from PCA was also used for galaxy clustering by \cite{Madgwick2003clustering}.
\cite{Yip_2004_iterpca} used PCA to classify a sample of SDSS galaxy spectra while \cite{Rogers2007} analysed a sample of SDSS early-type galaxies using PCA in order to infer differences in the average age of stellar population and star formation history. Following \cite{Slonim2001} on the information bottleneck method, \cite{Ferreras2022entropy} extracted the stellar population content of galaxy spectra by measuring the entropy of spectra in order to quantify the amount of information encoded. 
%Their study suggests that the $4000~\text{\r{A}}$ break and Balmer series are the most important sources of entropy variation.

Being a linear method, however, PCA can be limited when non-linear features are present. \cite{Yip_2004b_quasarclass} showed that $50$ PCA components are necessary to acceptably reconstruct a typical SDSS quasar spectrum exhibiting broad emission lines. Autoencoders (AEs) and Variational Autoencoders (VAEs) are non-linear deep-learning methods which have been successfully employed to address these limitations. In particular, \cite{Portillo2020vae} used a VAE on a sample of SDSS galaxy spectra. They showed that a VAE with six latent dimensions could satisfactorily reconstruct spectra whose original dimension was $1000$ and outperform PCA with the same number of components. The VAE latent space showed separation of classes and they demonstrated the interpretability of the latent space. Finally, they identified outliers by estimating the local density of samples in the latent space using the local outlier factor (LOF) algorithm and ranking spectra according to how isolated they were in the latent space. The top $10$ outliers identified were stars, spectra with the majority of the fluxes masked, low signal-to-noise ratio (S/N) spectra, bad instrument calibration and contamination by stellar light. 
\cite{Ichinohe2019} used a VAE to rank objects based on the reconstruction error. They approached this in the framework of novelty detection for a sample of simulated high-resolution X-ray spectra. 
%they use normalized chi-square statistic
Most recently, \cite{liang2023outlier} used an AE to compress galaxy spectra from the DESI Early Data Release (EDR, \citealt{https://doi.org/10.5281/zenodo.7964161}) into a redshift-invariant latent space, and a normalizing flow to detect outliers. Additionally, \cite{Scourfield_2023} used a VAE model to demonstrate their ability to denoise spectra.

Proximity-based approaches to anomaly detection in spectroscopic data include
\cite{Meusinger2012} who used self-organising maps \citep[SOM,][]{kohonen1982som} -- a clustering based algorithm
%that also makes use of non-linear dimensionality reduction
-- on a sample of SDSS quasar spectra. Unusual spectra were defined as those points to be found in the low density regions of representation created via the SOM. Outliers mainly included broad absorption lines, unusual red continua, weak emission lines and conspicuously strong iron emission lines. \cite{Fustes2013} also used SOM for outlier detection on a simulated sample of spectra based on SDSS observations.
\cite{Baron2017} used unsupervised random forests to identify outliers in SDSS spectra. Some interesting outliers identified include galaxies which host supernovae, high ionisation spectral lines, galaxies with unusual gas kinematics and galaxy-galaxy gravitational lenses.
%\cite{Skoda2020} active learning, used labelss, LAMOST spectra ref in lochhner, CNN
\cite{2022scio.confE...7S} used Uniform Manifold
Approximation and Projection (UMAP) as a technique to assess the data quality of DESI. They used UMAP to
project DESI nightly data into a 2-dimensional space where they were able to identify instrumentation outliers.

In this study, we follow a broadly similar approach to \cite{Portillo2020vae} and perform outlier detection using a VAE, but we apply it to astronomical spectra obtained from the DESI survey and utilise much larger datasets. Although \cite{liang2023outlier} probed a similar group of spectra for outliers, our VAE-driven approach is more straightforward as it does not seek to be redshift-invariant and instead makes direct use of the spectroscopic redshifts assigned by the DESI pipeline. In our work, we explore two approaches for identifying outliers. The first method uses the reconstruction error of the VAE as an anomaly score. The second method is proximity based, and is based on the position of spectra in the VAE latent space as anomalous spectra are expected to be isolated from the majority of the data. 
%To define how isolated a sample is we use density estimation.

We find that outlier spectra fall into a variety of categories, with the main distinction being between those which are anomalous because of physical features and those due to instrumental artefacts. Depending on the science question, certain types of outliers will be relevant while others will present a nuisance. We make use of Astronomaly \citep{Lochner2021astronomaly}, a software package which combines the input of a human expert with the processing power of machine learning to provide a curated list of outliers which are most relevant to the user.

This paper is organised as follows. An introduction to VAEs is provided in Section~\ref{sec:vaes}. In Section~\ref{sec:vae_implementation} we outline the specific VAE architecture we used, with a description of the DESI data and the preprocessing we applied in Section~\ref{sec:data}. The results are then split into five sections. In Section~\ref{sec:recon_acc} we examine the quality of the VAE reconstructions. In Section~\ref{sec:outlier_identification} we present the outliers obtained from the reconstruction error and latent space approaches. 
In Section~\ref{sec:artifial_anomalies} we inject synthetic anomalies into a representative galaxy spectrum to investigate how the VAE encodes atypical features and populates underdense regions of the latent space.
In Section~\ref{sec:astronomaly} we present the application of Astronomaly. In Section~\ref{sec:inter_latent_space} we explore the VAE latent space and the interpretability of the latent representations. Finally, in Section~\ref{sec:discussion} we discuss our results, along with limitations and improvements for future work.

%Three broad categories of anomaly detection methods are deviation based, proximity based and statistical. Deviation based anomaly detection uses the reconstruction error of a model trained on the data as the anomaly score. Anomalies will have a high reconstruction error as they have features that are atypical which the model is not able to reconstruct. Anomaly detection methods based on proximity assume that anomalous data are isolated from the majority of the data. Approaches to define how isolated a sample is include clustering, distance-based measurements and density estimation. Statistical anomaly detection makes use of parametric or non-parametric models to define a probability distribution assumed to describe the data. A sample is anomalous if it has a low probability of being generated by the model \citep{AnCho2015VaeAD, Reis2019Unsuprv}.

%amelie
%\cite{Ferreras2022entropy}

\section{Variational Autoencoders}\label{sec:vaes}
\begin{figure}%[b!]
    \centering
    \includegraphics[width=0.45\textwidth]{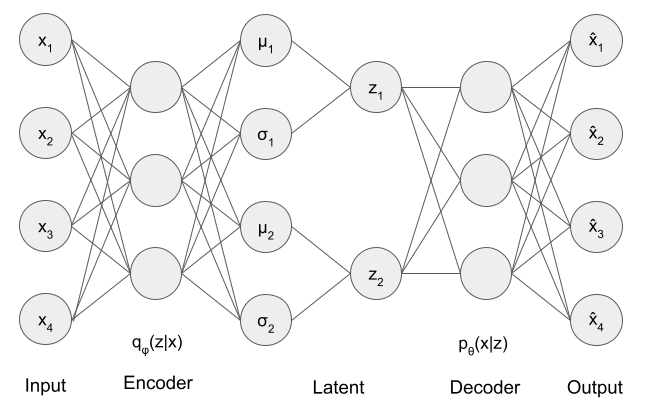}
    \caption{Example architecture of a Variational Autoencoder (VAE). The encoder is tasked with mapping the input, $x$ onto the lower dimensional latent space, $z$, as a normal distribution described by two vectors containing the latent mean $\mu$ and latent standard deviation $\sigma$. The decoder is tasked with decompressing samples from the latent representation and producing reconstructions, $\hat{x}$, of the original data.}
    \label{fig:vae}
\end{figure}

VAEs \citep{Kingma2013VAE} are probabilistic, generative models based on the encoder-decoder architecture. VAEs are tasked with learning efficient low-dimensional representations of the data in an unsupervised manner. Figure~\ref{fig:vae} depicts an example architecture of a VAE.

In a standard autoencoder \citep[AE,][]{KramerAE1991}, the encoder is typically a feedforward neural network and is tasked with mapping the input data, $x$, onto a low-dimensional hidden representation, $z$, called the latent space. As the dimension of $z$ is smaller than $x$, the encoder performs dimensionality reduction. This presents a bottleneck to the model and the encoder must learn an efficient compression that preserves as much of the information as possible by capturing the most meaningful factors of variation in the data. The decoder is also a feedforward neural network, typically mirroring the architecture of the encoder. Using just the latent space representations, it is tasked with reconstructing the original data. The reconstructed output, $\hat{x}$ is compared to the original data through the loss function which is typically given by the L$^2$ function, $||x - \hat{x}||$. This measures how far the reconstructed values are from the original ones in Euclidean terms. During training, this error is backpropagated, informing the weights update so as to achieve a lower loss value.
Hence, the AE attempts to find the best encoder-decoder pair so as to maximise the information encoded in the latent space and learn the underlying feature representations of the data.

While AEs are able to capture non-linearities in the data and learn powerful representations in low dimensions, the latent space they create is unregularised. While the encoder is able to usefully map similar inputs together in latent space forming clusters, this subspace may be discontinuous, with gaps between clusters that do not correspond to any data points. Using the decoder to try and generate from these gaps, may therefore result in unrealistic outputs. This limits the generative usefulness of the decoder. Additionally, the lack of a well-defined structure in the latent space may result in less reliable representations of input data, especially for outliers, leading to poor separation between normal and anomalous data.

The latent space of VAEs, on the other hand, is by design continuous and therefore allows for the decoder to randomly sample from any point in the space and generate meaningful outputs. Instead of encoding each input in a vector, the encoder produces a probability distribution over $z$, $~q_\phi(z|x)$, where $\phi$ represents the parameters of the encoder network. The encoded distribution is given by a normal distribution and hence can be described by two vectors containing the latent mean, $\mu$, and latent variance, $\sigma^2$. % (diagonal of the covariance).
A point from the latent space is then sampled from the encoded distribution according to

\begin{equation}
    z \sim q_\phi(z|x) = \mathcal{N}(\mu_{z|x}, \sigma^2_{z|x}).
\end{equation}
%%%
The sampled $z$ are then passed through the decoder. As the sampling procedure is stochastic, each time the same input is passed through the encoder during training, a slightly different sample in latent space will be propagated through the decoder. Hence the decoder will not learn just a single point in latent space that characterises the input, but it will learn that all points that fall within a small area in latent space defined by the variance of $q_\phi$ are representations of the input. 

The decoder also outputs a probability distribution, inferring $x$ from $z$. The probability distribution is given by $p_\theta (x|z)$, where $\theta$ are the parameters of the decoder network. The decoder distribution is chosen to reflect the type of input data. For example, in the case of continuous input data a normal distribution is used resulting in the outputs being given by

\begin{equation}
    \hat{x} \sim p_\theta (x|z) = \mathcal{N}(\mu_{x|z}, \sigma^2_{x|z}). 
\end{equation}

The VAE objective function (or loss function) is given by the evidence lower bound (ELBO) which is made up of two terms defined as follows:

\begin{equation}\label{eq:elbo}
    \begin{split}   
     \log p(x) \geq{}& \mathbb{E}_{z\sim q_\phi(z|x)} \left[ \log p_\theta(x|z) \right] - D_{KL}(q_\phi(z|x) || p(z)) \\
     \geq{}& \mathcal{L}(x)
    \end{split}
\end{equation}

\noindent where $\mathcal{L}(x)$ is known as ELBO as it represents the lower bound on $\log p(x)$.
%Details of how this is derived are given in APPENDIX \ref{Appendix A}.
The ELBO is tractable and thereby the VAE is able to apply backpropagation and optimise a lower bound on the likelihood of the data by finding the parameters of the VAE network ($\phi$ and $\theta$) that maximise the lower bound (or equivalently minimise the negative ELBO) \citep{RezendeVaeBackprop}. %This defines the objective function (or loss function) of the VAE. 

The first term of the ELBO is the expectation of $p_\theta(x|z)$, which is the decoder output, over the latent variables, $z$, sampled from the output of the encoder. By maximising this we are essentially maximising the likelihood of the observations given $z$ and hence this term represents the reconstruction likelihood which encourages the VAE to produce accurate reconstructions. 

The second term is a Kullback-Leibler divergence (KL divergence or $D_{KL}$, \citealt{MR0039968}) between $q_\phi(z|x)$ and $p(z)$. The KL divergence is a non-symmetric measure of similarity between two probability distributions.
%and hence can be used to measure how good of an approximate estimate $q_\phi(z|x)$ is to the actual distribution $p(z|x)$.
By maximising the ELBO, the second term is minimised which imposes that the approximate posterior distribution $q_\phi(z|x)$ is close to the prior distribution $p(z)$. The prior distribution is chosen to be a standard Gaussian $p(z)=\mathcal{N}(0,I)$ where $I$ is the identity matrix. As the prior on $z$ is diagonal, it encourages the latent variables to be independent and thereby encode different interpretable factors of variation. It is evident that this term acts as a regularisation term, enforcing the output of the encoder, and hence the latent variables $z$, to follow a standard Gaussian. This not only enables the generative process but also results in a regularised latent space that clearly separates normal from anomalous data points.

\section{VAE Implementation}\label{sec:vae_implementation}
We implemented a VAE using \texttt{tensorflow-probability} \citep{tensorflow2015-whitepaper}. The encoder consists of the input layer which is made up of $1000$ nodes, followed by $4$ hidden layers with dimensions $800, 600, 500, 300$. The architecture of the decoder mirrors that of the encoder. To illustrate this approach, the latent dimension is set to $10$. This is chosen based on effective reconstruction results demonstrated in the literature (e.g. \cite{Portillo2020vae}) but more rigorous testing is required to find the optimal number of latent dimensions (by carrying out a hyperparameter optimisation procedure for example) and to understand the effects of different dimension sizes \citep{Scourfield_2023}. Dropout is used in both the encoder and decoder to regularise the VAE and avoid overfitting. The dropout rate is set to $0.2$. The rectified linear unit (ReLU) activation function given by
\begin{equation}
    f(x) = 
    \begin{cases} 
    0 &\text{for $x<0$}\\
    x &\text{for $x\geq 0$}
    \end{cases}
\end{equation}
is adopted in all layers except for the output layers of the encoder and decoder where a linear activation function is used. We use a Gaussian distribution for the output of the decoder and thus the reconstruction loss term is given by the Gaussian log-likelihood. The first term in the loss function of the VAE (see eq.~\ref{eq:elbo}), is given by
\begin{equation}
    \mathbb{E}_{z} \left[\log p(x|z) \right] = \mathbb{E}_{z} \left[ \sum_i \frac{1}{2} -\log 2\pi \sigma_{x|z, i}^2 - \frac{(x_i - \mu_{x|z, i})^2}{2 \sigma_{x|z, i}^2}
    \right]
\end{equation}
where $i$ indicates the $i^{th}$ element of the vectors \citep{Yu2020VAE}, and log represents the natural logarithm.
The inverse variance associated with each flux measurement of each spectrum is used in the reconstruction loss, acting as a weighting term. This allows us to explicitly incorporate heteroscedastic uncertainties and element-wise masking. Bad pixels have their weight in the loss set to zero which is equivalent to infinite uncertainty.

We train the VAE using Adam optimiser with the starting learning rate set to $10^{-3}$. The learning rate is set to be reduced by a factor of $10$ if the loss function on the validation set does not improve for $5$ epochs. We use a batch size of $512$ and trained the model for $50$ epochs. For the purposes of this proof-of-concept study, it was felt sufficient to select VAE hyperparameters through a limited random search of the hyperparameter space. Optimising the VAE in a more rigorous manner, either by using a more extensive random search or a grid search \citep{Bergstra2012_randomsearch}, should reveal a truly optimised architecture which might be employed in future work.
%Training the VAE takes around $17Z\si{min}$ on Nersc shared cpu (ADM EPYC processor).
%random search https://www.jmlr.org/papers/volume13/bergstra12a/bergstra12a.pdf as we have no way of quantifying how well it is doing -we are not simply interested in in minimizing the MSE.

In VAEs using the standard evidence lower bound (ELBO) loss function (as in our implementation), anomaly detection arises naturally as the model learns to reproduce the dominant modes of the training data distribution. Data that deviate significantly from this learned distribution are reconstructed poorly or mapped to low-density regions in the latent space, thereby appearing as anomalies. Avoiding overfitting is crucial for anomaly detection: a model that perfectly reproduces all inputs, including atypical ones, would lose the ability to distinguish genuine outliers. We mitigate this through dropout regularisation, the KL divergence term, and inverse-variance weighting, all of which encourage the model to capture the general structure of the data rather than memorising individual examples. These design choices ensure that the anomaly signal reflects genuine departures from the learned spectral manifold. Future work could also explore the effectiveness of modified loss functions.

%\section{Methods}\label{sec:methods}
\section{DESI Data} \label{sec:data}
The Dark Energy Spectroscopic Instrument (DESI) is a robotic
%, fiber-fed, highly multiplexed 
spectroscopic surveyor that operates on the Mayall 4-meter telescope at Kitt Peak National Observatory (\citealt{DESI2022.KP1.Instr}). DESI, which can obtain simultaneous spectra of almost 5000 objects over an approximately 3-degree field (\citealt{DESI2016b.Instr, FocalPlane.Silber.2023, Corrector.Miller.2023}), is currently conducting a five-year survey of about a third of the sky. This campaign will obtain spectra for approximately 40 million galaxies and quasars (\citealt{DESI2016a.Science}).

The 
primary %RPN
goal of DESI is to determine the nature of dark energy through the most precise measurement of the expansion history of the universe ever obtained (\citealt{Snowmass2013.Levi}). DESI was designed to meet the definition of a Stage IV dark energy survey with only a 5-year observing campaign. Forecasts for DESI (\citealt{DESI2016b.Instr}) predict a factor of approximately five to ten improvement on the size of the error ellipse of the dark energy equation of state parameters $w_{0}$ and $w_{a}$ relative to previous Stage-III experiments.

The sheer scale of the DESI experiment necessitates multiple supporting software pipelines and products, which include significant imaging from the DESI Legacy Imaging Surveys (\citealt{LS.Overview.Dey.2019}), an extensive spectroscopic reduction pipeline (\citealt{Spectro.Pipeline.Guy.2023}), a template-fitting pipeline to derive classifications and redshifts for each targeted source (Redrock; Bailey et al, 2024 (in preparation), and for the special case of QSOs, \citealt{RedrockQSO.Brodzeller.2023}), and multiple pipelines with specific functions, e.g. to assign fibers to targets and to tile the survey and plan and optimise observations as the campaign progresses (\citealt{SurveyOps.Schlafly.2023}).

The dataset consists of DESI spectra from the Early Data Release (EDR) of the Bright Galaxy Survey (BGS, \citealt{Hahn2022DESIBGS, Myers2022DESITarget, Ruiz-Macias2021DESIBGS, 2024arXiv240403621J}). During the $5$-year DESI observing run, the BGS will observe $10$ million galaxies over the redshift range $0 < z < 0.6$. The BGS targets galaxies with a limit in the r-band magnitude of $r<19.5$ (BGS Bright), a secondary lower priority sample defined by the magnitude range $19.5 < r < 20.175$ (BGS Faint) and a small low-z quasar sample (BGS AGN). 
%% Chose the BGS sample because it is the largest one and even after snr cuts etc we would still be left with a large sample. Also it is the largest sample at low redshift https://desi.lbl.gov/trac/wiki/Pipeline/Releases/Fuji

The redshift and spectral classifications are obtained from the DESI redshift pipeline, named \texttt{redrock} \footnote{https://github.com/desihub/redrock} which is a template-fitting code, tasked with finding an optimal fit to a spectrum via linear combinations (PCA) of the template components derived from galaxy models \citep{Spectro.Pipeline.Guy.2023}. The fitting is performed over a specified redshift range for three template classes: “star”, “galaxy”, and “quasar”. 

Using the redrock information, we select objects with redshifts in the range $0 \leq z \leq 0.3$, and convert the observed wavelength vector of each object to the rest-frame. The analysis is performed over the rest-wavelength range of $3600~\text{\r{A}}$ to $7556~\text{\r{A}}$, which is fully sampled by all the objects in the chosen wavelength range. The analysis could be repeated over higher redshift intervals. We perform anomaly detection in the rest-frame so that spectral features with the same physical origin align in wavelength. This allows the VAE to learn variations in intrinsic spectral information - line ratios, continuum shapes, and unusual features - rather than trivial wavelength shifts caused by redshift. Training in the observed frame would force the model to waste capacity learning that otherwise identical spectra differ only by a translation. A natural by-product of this choice is that we expect spectra with incorrect redshift estimates to be flagged as anomalous since their features will be misaligned relative to the training distribution, providing a useful means to identify potential redshift misclassifications. Alternative observed-frame anomaly detection strategies could employ convolutional VAEs that explicitly learn redshift-invariant representations. We leave exploration of such shift-invariant models to future work.

%We use the redshift provided by \texttt{redrock} to shift the spectra to their rest-frame %according to eq.~\ref{eq:z_lambda_a},
%so that the spectral lines from all spectra align at the same wavelength bin. As the spectra in the dataset all have different redshifts but their observed wavelength range is the same, however, shifting the spectra to the rest-frame wavelength means there is now a misalignment in the feature space of the dataset.
%Wavelength bins for which we do not have information from all the spectra are removed. To make sure we are not throwing away large parts of a spectrum, we limit the dataset to $0 \leq z \leq 0.3$ corresponding to a common 
%wavelength range of $3600~\text{\r{A}}$ to $7556~\text{\r{A}}$ in the rest-frame. The analysis can then be repeated for higher redshift bins. 
%The original dimension of spectra is $\sim 7780$ which means that each spectrum is described by $\sim 7780$ flux measurements. 
We resample the spectra to $1000$ wavelength bins ensuring that the total integrated flux remains consistent before and after resampling. While this leads to some small-scale information being lost as the DESI spectral resolution is much higher, the resampling is done to reduce the computational cost. This resampling corresponds to a resolution of $4~\text{\r{A}}$ which is sufficient to avoid blending between important spectral lines such as \halam\ and \niidoub. Note that we have opted for a simple resampling procedure which did not account for covariances between resampled bins and therefore this may lead to a slight underestimation of the uncertainties. Additionally, before the resampling procedure, pixels with missing values, zero flux or zero inverse variance are removed and a simple linear interpolation is used to fill in the values.

Each resampled spectrum has an indicator mask vector that denotes bad pixels e.g. pixels that are affected by CCD (charged-coupled device in the spectrograph cameras) defects or cosmic rays. Bad pixels are removed and iterative PCA \citep{Yip_2004_iterpca} is used to infill the missing values. The dataset does not suffer from heavily masked spectra as only $0.01\%$ of the spectra have more than $10\%$ of the pixels masked.
The spectra are normalised individually to have unit norm.
The median pixel signal-to-noise ratio (S/N) is calculated for each spectrum. We only keep the spectra that have a median pixel S/N above $5$.
%[WHAT METHOD IS BEING USED HERE TO CALCULATE SNR?]

The resulting dataset consists of $\sim208,000$ spectra, of which $\sim156,000$ form the training set and the rest form the validation set. In future work, the study presented here can be scaled to a larger dataset by making use of more spectra, which will also have a greater potential to lead to a novel object being identified.
Using the spectral classifications provided by \texttt{redrock}, the dataset is composed of around $97.8\%$ galaxy spectra, $0.7 \%$ quasar spectra and $1.5\%$ star spectra. Note however that these are not true labels and while \texttt{redrock} performs well on galaxies and stars it misclassified $\sim 10-15 \%$ of true quasars, in particular low-redshift ones \citep{Alexander_2023}.
%We note that we do not include the object IDs of spectra when presenting results, as the DESI data are not published yet.

%%%%%% ELLIPTICAL
\begin{figure*}
    \centering
    \includegraphics[width=1\textwidth]{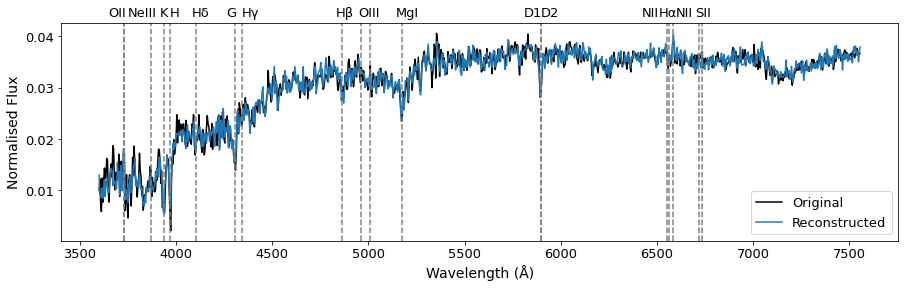}
    \caption{Original (black) and reconstructed (blue) spectrum based on 10 latent variables (in this and and subsequent figures) of an \textbf{elliptical galaxy} from the validation set. The median pixel S/N of the galaxy is $33$. The spectrum is in the rest frame. The dashed grey lines indicate typical spectral lines labelled at the top of the plot. Target ID: 39627836461419062.} 
    \label{fig:recon_spectrum_elliptical}
%\end{figure}

%\begin{figure}%[h!]
\subfloat{
  \includegraphics[clip,width=0.87\textwidth]{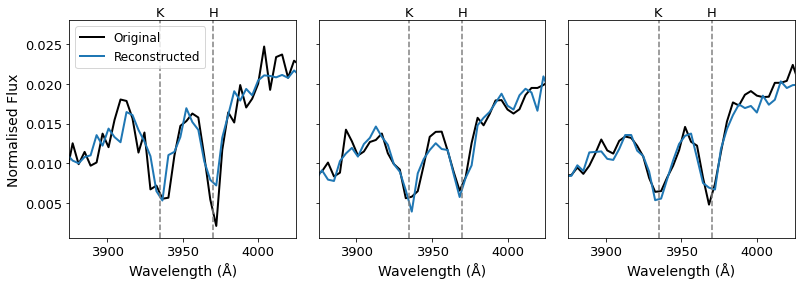}}

\subfloat{
  \includegraphics[clip,width=0.87\textwidth]{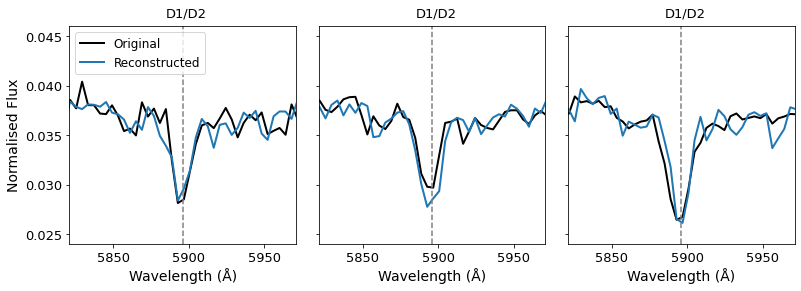}}
\caption{Original (black) and reconstructed (blue) spectra of \textbf{three elliptical galaxies}. The top row shows the VAE reconstruction of the \caiihk\ lines and the bottom row the \dlines. Target IDs left to right: 39627836461419062, 39633321180793652 and 39628435806491872.}
\label{fig:recon_zoomin_spectrum_elliptical}
\end{figure*}

%%%%%%% EMISSION
\begin{figure*}[H]
    \centering
    \includegraphics[width=1\textwidth]{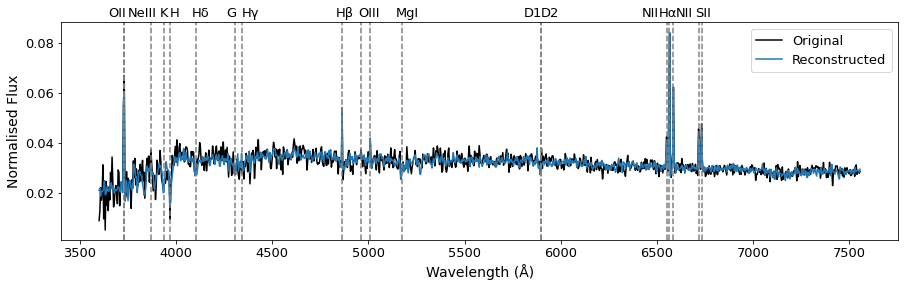}
    \caption{Original (black) and reconstructed (blue) spectrum of an \textbf{emission-line galaxy} from the validation set. The median pixel S/N of the galaxy is $14$. The spectrum is in the rest frame. The dashed grey lines indicate typical spectral lines labelled at the top of the plot. Target ID: 39627974831507323.} 
    \label{fig:recon_spectrum_emission}
%\end{figure}

%\begin{figure}%[h!]
\subfloat{
  \includegraphics[clip,width=0.87\textwidth]{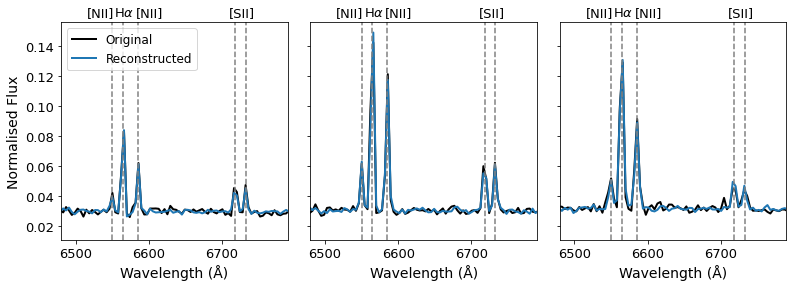}}

\subfloat{
  \includegraphics[clip,width=0.87\textwidth]{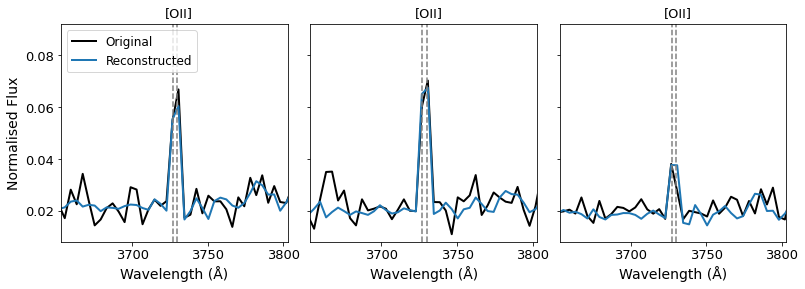}}
\caption{Original (black) and reconstructed (blue) spectra of \textbf{three emission-line galaxies}. The top row shows the VAE reconstruction of the \halam, \niidoub\ and \siilam\ lines and the bottom row the \oiilam\ lines. Target IDs left to right: 39627974831507323, 39633342945036446 and 39628521886189863.}
\label{fig:recon_zoomin_spectrum_emission}
\end{figure*}

\section{Results}
\label{sec:results}
\subsection{Reconstruction Accuracy}
\label{sec:recon_acc}
First let us consider how well the VAE is able to reconstruct the spectra  of various objects. 
Even though the VAE was not optimised to achieve the lowest possible reconstruction error, it is still important to assess whether it is able to reconstruct the spectra to a reasonable degree. 
To quantify the performance of the VAE on the task of reconstruction, we calculate the weighted mean squared error (MSE) between the original and reconstructed spectrum. The weighted MSE for the $i^{th}$ spectrum is defined as

\begin{equation}\label{weightedMSE}
    \text{weighted MSE}_i = \frac{1}{M} \sum_{m=1}^{M} \bm{w}_i \cdot (\bm{x}_i - \bm{\hat{x}}_i)^2
\end{equation}
where $\bm{x}_i$ is the original spectrum, $\bm{\hat{x}}_i$ is the the reconstructed spectrum and $\bm{w}_i=\frac{1}{\bm{\sigma}_i^2}$ where $\boldsymbol{\sigma_i}$ is the noise estimate from the data. All three vectors have dimensions $M=1000$ corresponding to the $1000$ flux measurements.
The mean weighted MSE on the validation set, as defined in Eq.\ref{weightedMSE}, is $1.12$. Broken down to the three classes, galaxies have the lowest average MSE at $1.09$, stars have a MSE of $1.87$ and quasars $2.57$. Since our dataset is mostly composed of galaxies the VAE was able to see many examples of galaxy spectra and hence learn to reconstruct them well. The VAE however, struggles to reconstruct stars and quasars in comparison to galaxies, as these are rare in the dataset. The VAE hasn't seen enough star and quasar spectra examples in order to learn the underlying correlations and the most efficient representation in the latent space. This demonstrates that the reconstruction error is an effective metric for identifying anomalies. 
%Spectra that have good reconstructions are described well by the underlying distribution of the majority of the dataset and hence the VAE is able to efficiently describe them using the latent dimensions. Spectra that have not been reconstructed well are expected to contain features that are atypical and differ from the majority of the dataset and hence their latent space representation is not as good.

%%%%%
%%%%%%%% AGNs/Seyferts
\begin{figure*}
    \centering
    \includegraphics[width=1\textwidth]{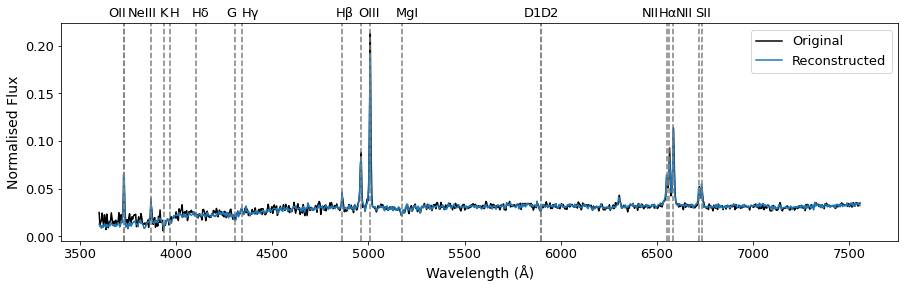}
    \caption{Original (black) and reconstructed (blue) spectrum of an \textbf{AGN} from the validation set. The median pixel S/N of the galaxy is $17$. The spectrum is in the rest frame. The dashed grey lines indicate typical spectral lines labelled at the top of the plot. Target ID: 39633118658822194.} 
    \label{fig:recon_spectrum_seyf2}
%\end{figure}

%\begin{figure}%[h!]
\subfloat{
  \includegraphics[clip,width=0.87\textwidth]{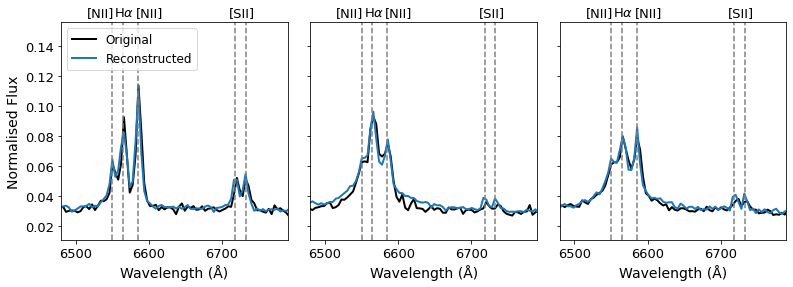}}

\subfloat{
  \includegraphics[clip,width=0.87\textwidth]{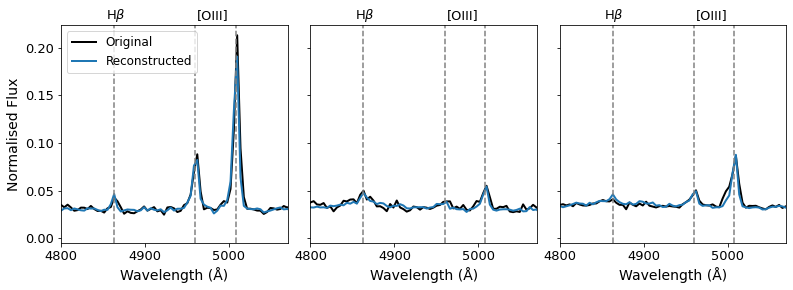}}
\caption{Original (black) and reconstructed (blue) spectra of \textbf{three AGNs}. The top row shows the VAE reconstruction of the \halam, \niidoub\ and \siilam\ lines and the bottom row the \hblam\ and \oiiilam\ lines. Target IDs left to right: 39633118658822194, 39628133594305704 and 39627794468045508.}
\label{fig:recon_zoomin_spectrum_seyf2+1}
\end{figure*}

%%%%%%%%%%% STAR
\begin{figure*}
    \centering
    \includegraphics[width=1\textwidth]{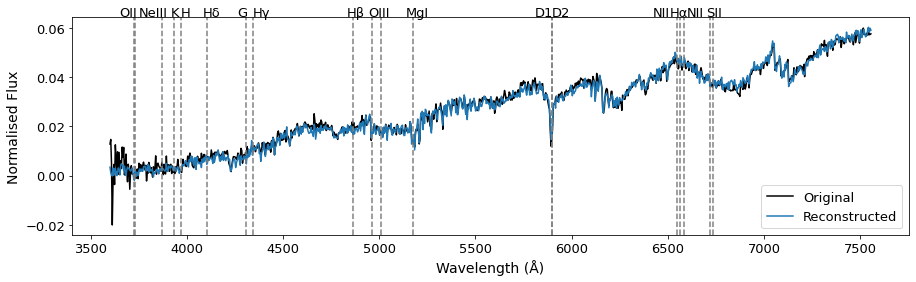}
    \caption{Original (black) and reconstructed (blue) spectrum of a \textbf{star} (M type) from the validation set. The median pixel S/N of the spectrum is $27$. The spectrum is in the rest frame. The dashed grey lines indicate typical spectral lines labelled at the top of the plot. Target ID: 39627582181739826.}
    \label{fig:recon_spectrum_star}
\end{figure*}

\subsubsection{Galaxy Spectra Reconstructions}
Figure~\ref{fig:recon_spectrum_elliptical} shows the original and reconstructed spectrum of an quiescent galaxy obtained from the validation set. Note that in this figure and subsequent figures we plot the original and reconstructed spectra at the resampled wavelength of $1000$ bins. The median pixel S/N of the spectrum is $33$ and the weighted MSE is $1.19$. The VAE is able to match the continuum and absorption lines of the spectrum well. The $4000~\text{\r{A}}$ break is present and captured well by the VAE. 
%This feature is enhanced in galaxies with old metal-rich stellar populations. The lack of hot, blue stars and the absorption of high energy photons by metals in the atmosphere of cool stars lead to a drop in flux in the blue part of the spectrum. 
In Figure~\ref{fig:recon_zoomin_spectrum_elliptical} we zoom in on regions of the spectrum that contain spectral lines. We show the \caiihk\ which are prominent absorption lines in the spectra of solar-type and cooler stars due to singly ionised calcium and the \dlines\ which occur in the yellow region of the spectrum due to sodium.
The VAE is able to reconstruct well the absorption lines, capturing well both the width and amplitude but in some cases the VAE reconstruction undershoots or overshoots slightly, failing to exactly meet the amplitude of the spectral line. Getting the best possible reconstruction accuracy is beyond the scope of this work. However, this could be achieved by using higher-resolution spectra, optimising the VAE hyperparameters, and training on a cleaned homogeneous dataset (e.g., galaxy-only or quasar-only spectra), rather than the mixed dataset used in this study.

Figure~\ref{fig:recon_spectrum_emission} shows the original and reconstructed spectrum of an emission-line galaxy with a median pixel S/N of $14$. The VAE is able to reconstruct it well with a weighted MSE of $0.84$. The spectrum shows strong emission lines in \ha, \nii, \sii, \hb\ and \oii\ indicating that star formation is occurring.
In Figure~\ref{fig:recon_zoomin_spectrum_emission} we zoom in on two regions of interest to observe more closely the \ha, \nii, \siilam, \hblam and \oiilam\ emission lines.
The VAE is able to reconstruct these very well, accurately mapping the width and height of the peaks and does not show any blending of the lines that are closely located.

\subsubsection{AGN Spectra Reconstructions}

Figure~\ref{fig:recon_spectrum_seyf2} shows the reconstructed spectrum of an active galactic nucleus (AGN), selected according to the \texttt{redrock} SPECTYPE of QSO,
with median pixel S/N of $17$. The VAE is able to reconstruct the AGN spectrum with a weighted MSE of $1.68$. In Figure~\ref{fig:recon_zoomin_spectrum_seyf2+1} we zoom in on regions of interest. We observe the significant broadening of emission lines especially around the \ha\ region. Due to its non-linear nature, the VAE is able to capture well the width of the emission lines. \cite{Portillo2020vae} have also shown that VAEs successfully reconstruct broad emission lines overcoming the limitations of PCA which requires more components, compared to a VAE, to adequately reconstruct the width of broad lines.

\subsubsection{Star Spectrum Reconstruction}
The reconstructed spectrum of a star is shown in Figure~\ref{fig:recon_spectrum_star}. The spectrum has a median pixel S/N of $27$ and the VAE is able to reconstruct it with a weighted MSE of $1.8$. Even though there were comparatively few stars within the training set, the VAE still manages to reconstruct the stellar spectra with only slightly higher mean error than the galaxies and quasars.

\subsection{Outlier Identification}\label{sec:outlier_identification}

VAEs are increasingly adopted in astronomy for anomaly detection on diverse data types. \citet{Portillo2020vae} applied a VAE to SDSS spectra, showing that its latent space captures non-linear structure and naturally highlights outliers. \citet{Villar21} used a recurrent VAE to identify rare extragalactic transients from a dataset of simulated light curves, while
%, while \citet{Tohill2024} applied VAEs to JWST imaging for unsupervised galaxy morphology classification. 
\citet{Xiang} modelled inactive stellar spectra with a VAE to detect abnormal spectra (magnetically active and lithium-rich stars). \citet{gagliano} introduced a physics-informed convolutional VAE that enables both rapid galaxy parameter inference and discovery of rare populations. These studies demonstrate the versatility of VAEs as unsupervised tools for uncovering anomalies in large astronomical datasets.

In VAEs, outliers are expected to be badly reconstructed leading to high MSE or be placed in an isolated region in the latent space, i.e. the lower-dimensional representation produced by the VAE encoder.
%The studies performed in the previous sections (\ref{sec:recon_acc} and \ref{sec:inter_latent_space}), where we studied the reconstruction capabilities of the VAE and the latent space, have already indicated that the reconstruction error and location in latent space are good indicators for anomaly detection.
An anomalous spectrum may have both a high MSE and be isolated in latent space, but this is not necessarily true for all outliers as the latent space position does not only depend on the reconstruction error but depends on the entire loss function of the VAE. For example, an outlier, might be reconstructed badly because the VAE placed it together with a group of typical spectra in the latent space. In this case, the VAE will attempt to reconstruct the anomalous spectrum as a normal one. This will lead to a bad reconstruction and a high MSE, but the spectrum will not be located in an isolated region. The spectrum will have a similar local density to its neighbours (as it is in a populated area) but it will have a much higher MSE than its neighbours. 
A case where a spectrum is located in an isolated region of the latent space but does not have a high MSE might be a low S/N spectrum where the pixel errors are high. Since the MSE is weighted by the inverse variance this will lead to a lower MSE.
In this study we explore both approaches for identifying outliers and compare the results.

\subsubsection{Reconstruction Error Approach}
\begin{figure}%[t!]
     \centering
     \includegraphics[width=0.45\textwidth]{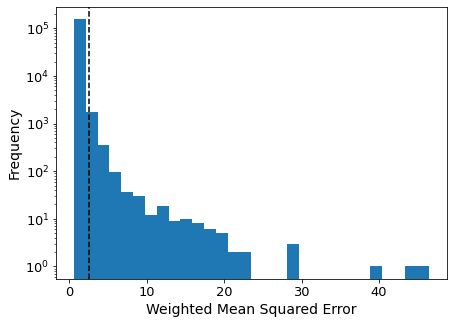}
     \caption{Histogram of the weighted MSE for all the spectra in the training set. The vertical dashed line represents the $1\%$ threshold and corresponds to an MSE value of $2.55$. Spectra with a MSE higher than this threshold constitute the top $1\%$ outliers in the dataset. Note the y scale is logarithmic.} 
     \label{fig:mse_hist}
\end{figure}

First, we explore the reconstruction error approach for detecting outliers. For each spectrum in the training set we calculate the weighted MSE between the original and reconstructed spectrum. For the purposes of this work we have chosen to identify the most extreme $1\%$ of points as outliers for investigation. Figure~\ref{fig:mse_hist} shows a histogram of the weighted MSE from all spectra. As expected, most spectra have a low MSE. The $1\%$ threshold is indicated by the vertical dashed line and corresponds to a MSE of $2.55$. This threshold suggests that spectra with an MSE higher than $2.55$ are outliers while the rest are inliers, resulting in $1,560$ outliers and $154,392$ inliers. It is interesting to note that the average MSE of quasars reported earlier in Section~\ref{sec:recon_acc} is 2.57 which falls within the $1\%$ threshold. This is because quasars are rare in the training sample and as such they are identified as outliers by the VAE. 
We selected the threshold at $1\%$ to focus on the most extreme cases. Although this threshold is arbitrary, it follows standard practice in anomaly-detection studies (e.g. \citealt{han2022adbenchanomalydetectionbenchmark}) and was guided by inspection of the score distribution to capture the tail of the population. The optimal threshold level depends on several factors such as dataset size, signal-to-noise ratio, and preprocessing quality, and represents a trade-off between sensitivity (true positive rate) and specificity (true negative rate). In the absence of ground-truth labels, determining an appropriate threshold ultimately requires domain expertise and depends on the scientific objective. For a more adaptive framework that does not require a fixed cutoff, we refer the reader to the Astronomaly method described in Section~\ref{sec:astronomaly}.

\begin{figure*}%[t!]
     \centering
     \includegraphics[width=1\textwidth]{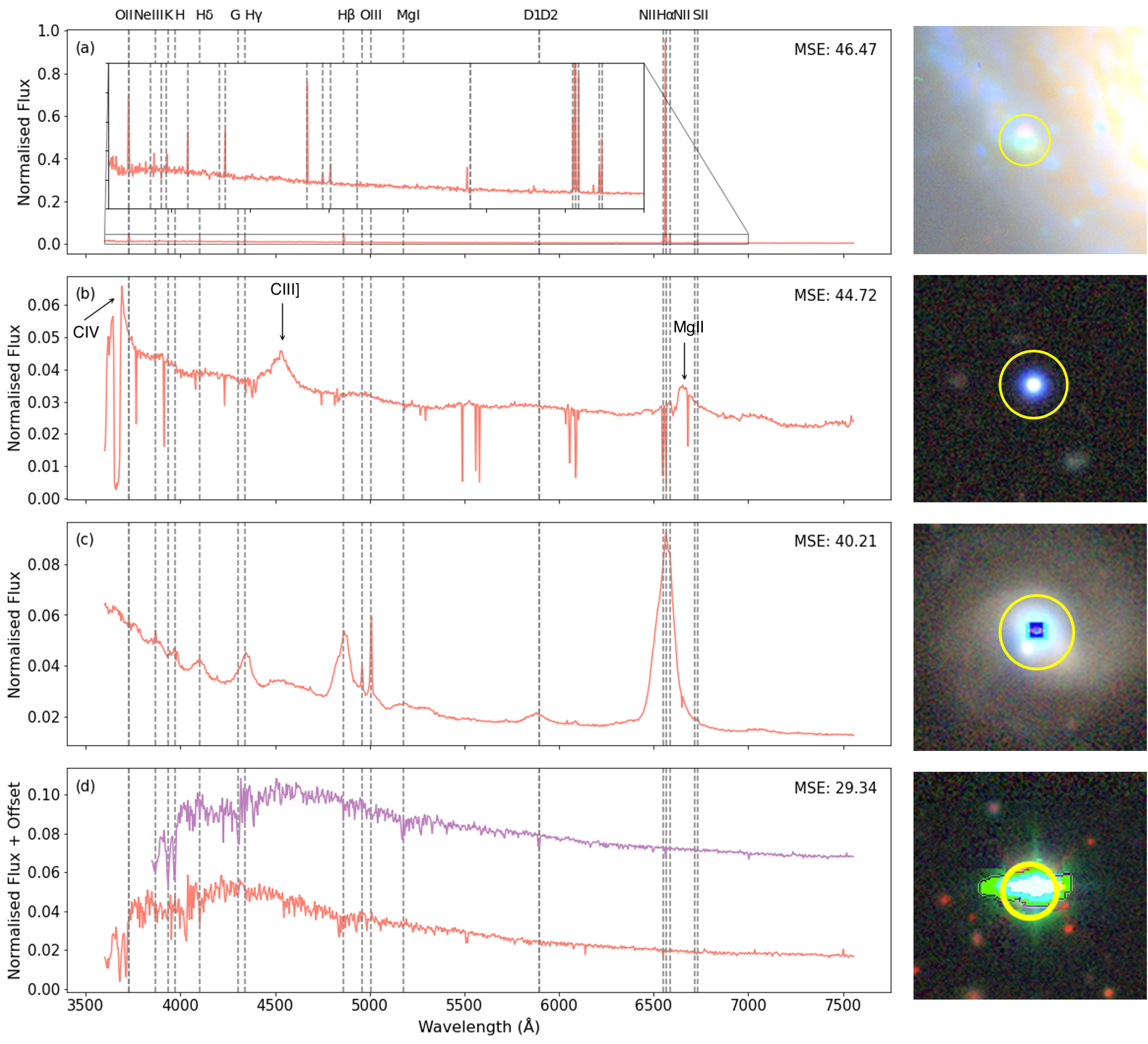}
     \caption{Top four outliers according to the highest reconstruction error of the VAE. The MSE of each spectrum is given in the upper right corner of each spectrum panel. The spectrum in panel (a) exhibits extreme \ha\ emission strength. The inset provides a magnification of difficult to see emission lines of the spectrum. The spectrum in (b) is of a high redshift quasar whose redshift was underestimated by \texttt{redrock}. We demonstrate the correct feature labels in the annotations. The spectrum in (c) is of an AGN exhibiting broad emission lines. The spectrum in (d) plotted in coral is of a star whose redshift was overestimated. The correct rest-frame spectrum is overplotted in purple. The object corresponding to each spectrum is shown in the images on the right hand side. The yellow circle is a marker indicating the objects of interest of various angular sizes. The images were obtained from the Legacy Survey sky viewer (data release 9). Target IDs from top to bottom: 39627763589581458, 39627817943567887, 39627818572710520 and 39633152305530204.} %as the original redshift, as given by \texttt{redrock}, was much higher than the true redshift of the object.}
     \label{fig:mse_outliers}
\end{figure*}

In Figure~\ref{fig:mse_outliers} we plot the top four outliers ranked according to highest MSE from top to bottom, together with images retrieved from the Legacy Survey sky viewer (data release 9).\footnote{https://datalab.noirlab.edu/ls/ls.php -- images maybe retrieved for objects with a given Target ID, RA and dec.} 
The spectrum with the highest reconstruction error is shown in Figure~\ref{fig:mse_outliers} (a) and has a MSE of $46.47$.  It has an \ha\ emission line of extreme strength. The flux ratio  of \ha\ to any other emission line present is extremely large, which is uncommon in the dataset. The spectrum exhibits emission lines in the Balmer series, \oii\ and \nii\ primarily, with weaker emission lines in \oiiilam, \heilam\, and \sii\ (shown in the magnified inset). This suggests intense star formation in a low-metallicity starbursting region as part of a galaxy. This results in lines which are very strong and also narrow.
%dominated by \hii\ regions ionised by newly formed massive hot stars.
Such extreme emission line ratios are atypical in the dataset and thereby this spectrum is flagged as an outlier.
From the image next to the spectrum we can observe that is not a galaxy but rather a fiber that has been placed on a particularly bright region inside of a very extended nearby galaxy. It is promising that the VAE is able to identify this as an outlier as it can be used as a tool for filtering bad spectra and providing a clean sample of galaxies.
 
Figure~\ref{fig:mse_outliers} (b) shows the spectrum of a quasar. \texttt{redrock} classifies this spectrum as a galaxy with an estimated redshift of $0.257$. However the spectrum corresponds to a quasar found at a much higher redshift, around $2$. Applying the correct redshift would set the rest-frame spectrum in the ultraviolet part of the EM spectrum. The correct rest-frame spectrum corresponding to this object is not plotted here as it is well outside the wavelength range of this study, but we annotate three main features (\mgii, \ciii and \civ) on the plot to demonstrate the correct labelling of features and aid with visualisation.
The VAE is unfamiliar with these out-of-place features and hence does not reconstruct them well
%. It instead tries to reconstruct this spectrum as an elliptical galaxy 
resulting in a MSE of $44.72$. This example shows, that using the VAE we are able to identify spectra that have been assigned a wrong class and redshift by the DESI spectroscopic pipeline. There has been much work in the AGN community to amend these types of errors using traditional methods, so this is an example of how VAE-based methods might be able to complement these efforts.

Figure~\ref{fig:mse_outliers} (c) shows the spectrum of an AGN with both broad and narrow features, suggesting a Seyfert 1 galaxy. As discussed earlier, AGN spectra constitute less than $1\%$ of the dataset with Seyfert 1 being a fraction of that. As these are rare in the dataset, they are picked up as anomalous.

Conversely to the case of the outlier spectrum in  Figure~\ref{fig:mse_outliers} (b) where \texttt{redrock} underestimated the redshift of the source, in (d) the redshift is overestimated. The spectrum exhibits the shape of a black body with multiple absorption lines suggesting the source as a star. However the spectrum was labelled as a galaxy by \texttt{redrock} with a redshift of $0.0692$. The true redshift of the star is smaller ($z \sim 0$). The correct rest-frame spectrum is shown in purple where the two strong absorption lines at the blue end of the spectrum now match the \caii\ K and H lines and the $4000~\text{\r{A}}$ break is matched to the correct wavelength. 

Inspecting by eye the $100$ highest MSE spectra, the outliers fall into one or more of the following broad categories: extreme emission line(s), stars, AGNs, discontinuities arising from miscalibration between neighbouring cameras, wrong assigned redshift, cataclysmic variables, and very broad emission lines (with the full width at half maximum spanning several hundred Angstroms). The top $100$ outliers have a median MSE of $13$ and median S/N of $42$.

\subsubsection{Local Outlier Factor Approach}
\begin{figure}%[t!]
     \centering
     \includegraphics[width=0.45\textwidth]{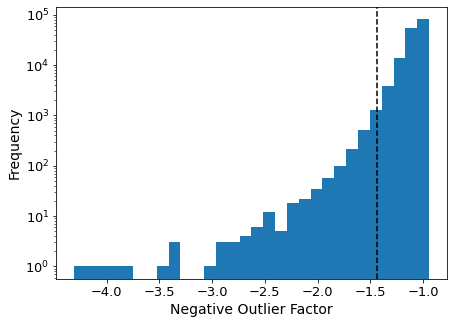}
     \caption{Histogram of the negative outlier factor (NOF) for all the spectra in the training set calculated using the local outlier factor (LOF) algorithm. Spectra with scores significantly lower than -1 are considered outliers. The vertical dashed line represents the $1\%$ threshold and corresponds to a NOF of $-1.44$. Spectra with a NOF score lower than this threshold constitute the top $1\%$ outliers in the dataset. Note the y scale is logarithmic.} 
     \label{fig:nof_hist}
\end{figure}

 In Section \ref{sec:inter_latent_space} we shall explore how the VAE latent space is organised and how it arranges data points of different object type. For now, we use the VAE latent space to find outliers, specifically by identifying those points which are in relatively isolated areas of the latent space. To do this, we use the local outlier factor (LOF) algorithm \citep{Breunig2000_LOF}. This unsupervised algorithm estimates the local density of a given spectrum in the latent space and compares it to the local density of its $k$ nearest neighbours. A spectrum is considered an outlier if its local density in the VAE latent space is substantially lower than that of its neighbours. Here we use the LOF algorithm with $k=20$ nearest neighbours. The LOF algorithm computes the negative outlier factor (NOF) for each spectrum. Inliers tend to have a NOF close to $-1$ while outliers tend to have more negative NOF scores. 
Figure~\ref{fig:nof_hist} shows a histogram of the NOF of all training samples. Most spectra have a NOF close to $-1$ indicating that these are inliers. The $1\%$ threshold, shown by the vertical dashed line, corresponds to a NOF of $-1.44$ suggesting that spectra with a NOF lower than $-1.44$ are considered as outliers.

\begin{figure*}%[t!]
     \centering
     \includegraphics[width=1\textwidth]{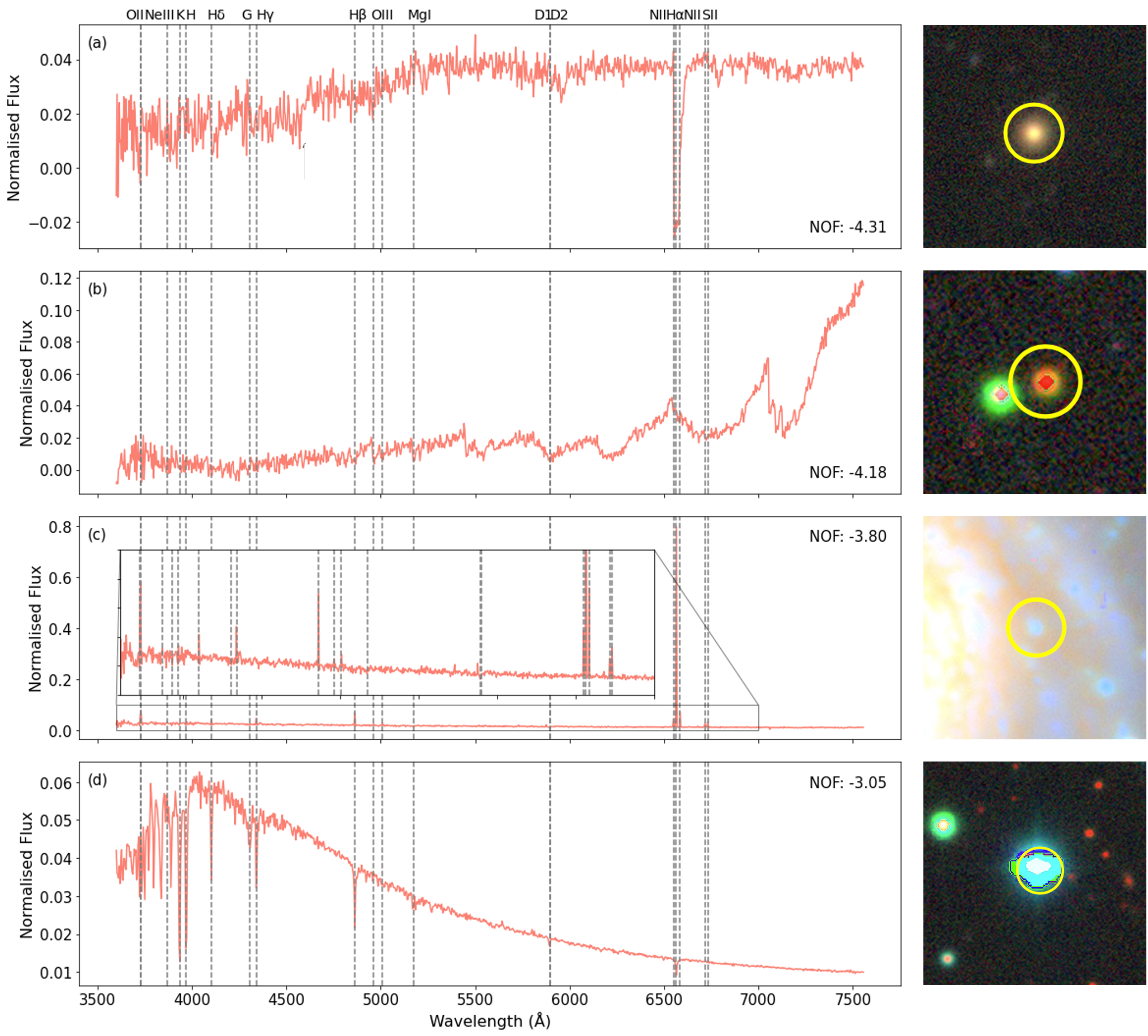}
     \caption{Four unique outliers according lowest negative outlier factor. The NOF of each spectrum is found at the upper or lower right corner of each panel. (a) has the lowest NOF, (b) the second lowest, (c) the fifth and (d) the tenth. Similar spectra have been skipped in order to avoid repeatability and present a greater variety of outliers. The galaxy spectrum in panel (a) exhibits a large dip in flux likely due to a bad sky subtraction. The arrow in (a) points to what we suspect is the misplaced $4000~\text{\r{A}}$ break, suggesting that the redshift of the galaxy is underestimated. The spectrum in (b) is of a M star with particularly high flux in the red part of the spectrum. The spectrum in (c) exhibits an extreme emission strength in \ha. The inset in (c) provides a magnification of the weaker emission lines of the spectrum. (d) shows the spectrum of a star. The object corresponding to each spectrum is shown in the images on the right hand side. The yellow circle is a marker indicating the objects of interest of various angular sizes. The images were obtained from the Legacy Survey sky viewer (data release 9).} 
     \label{fig:nof_outliers}
\end{figure*}

In Figure~\ref{fig:nof_outliers} we plot four outliers of different type according to lowest NOF. Figure~\ref{fig:nof_outliers} (a) and (b) have the lowest and second lowest NOF, (c) has the fifth and (d) the tenth lowest score. In order to present a greater variety of outliers, similar spectra have been skipped in this selection. The spectrum in Figure~\ref{fig:nof_outliers} (a) has a NOF of $-4.13$. This galaxy spectrum has a large dip in the flux with negative flux values. This is probably attributed to a bad sky subtraction or bad pixels which were not caught by the DESI spectroscopic pipeline. We also suspect that the assigned redshift is underestimated. 
The arrow points to what we suspect is the misplaced $4000~\text{\r{A}}$ break.

The spectrum in Figure~\ref{fig:nof_outliers} (b) is of an M star with particularly high flux in the red part of the spectrum.  The spectrum in Figure~\ref{fig:nof_outliers} (d) is also of a star.

The spectrum in Figure~\ref{fig:nof_outliers} (c) shows extreme emission in \ha\ which is similar to the top outlier from the MSE approach. The inset provides a magnification of the weaker emission lines. This spectrum has a MSE of $8.46$ corresponding to the $103$rd highest MSE. In fact, the spectrum with the highest MSE (Figure~\ref{fig:mse_outliers} (a)) has a NOF of $-3.38$ which corresponds to the seventh lowest NOF score. 
All $4$ outliers presented here obtained using the LOF approach are also in the top $1\%$ of outliers with the highest MSE but their relative ranking is lower.

Inspecting by eye the top $100$ outliers resulting from the LOF approach, we observe that these fall into one or more of the following broad categories: bad spectral regions which were not flagged by the mask vector and hence were not treated in the preprocessing, low S/N, extreme emission line(s), stars, AGNs, discontinuities due to bad calibration and wrong redshift. The top $100$ outliers have a median NOF of $-2.25$ and a median S/N of $7$. 

While the top $100$ outliers of the two methods share some categories, their median S/N differs markedly, indicating that noise influences a spectrum’s position in latent space. Lower S/N spectra have smaller weights, so deviations between the reconstructed and original spectra are dampened, keeping the weighted MSE low. However, noise is still propagated through the encoder, introducing uncertainty in latent-space position and making the LOF approach more sensitive to S/N. 
The overlap between the top 100 anomalies identified by the MSE and LOF methods is approximately $10\%$, showing that the two approaches respond to distinct types of outliers. Spectra with large reconstruction errors (high MSE) correspond to off-manifold anomalies, where the VAE fails to reproduce inputs lying outside the learned manifold, while the LOF method can highlight more on-manifold anomalies that occupy low-density regions. This complementarity is evidenced by the low overlap and also highlighted by \cite{pegasus}, which emphasises the benefits of combining on- and off-manifold approaches to achieve broader anomaly coverage. As this study serves as a proof of concept on a limited sample, a comprehensive classification of anomaly types will be pursued in future work on a larger DESI dataset.

\subsection{Injecting Synthetic Anomalies}\label{sec:artifial_anomalies}
\begin{figure}
\centering
\includegraphics[width=0.48\textwidth]{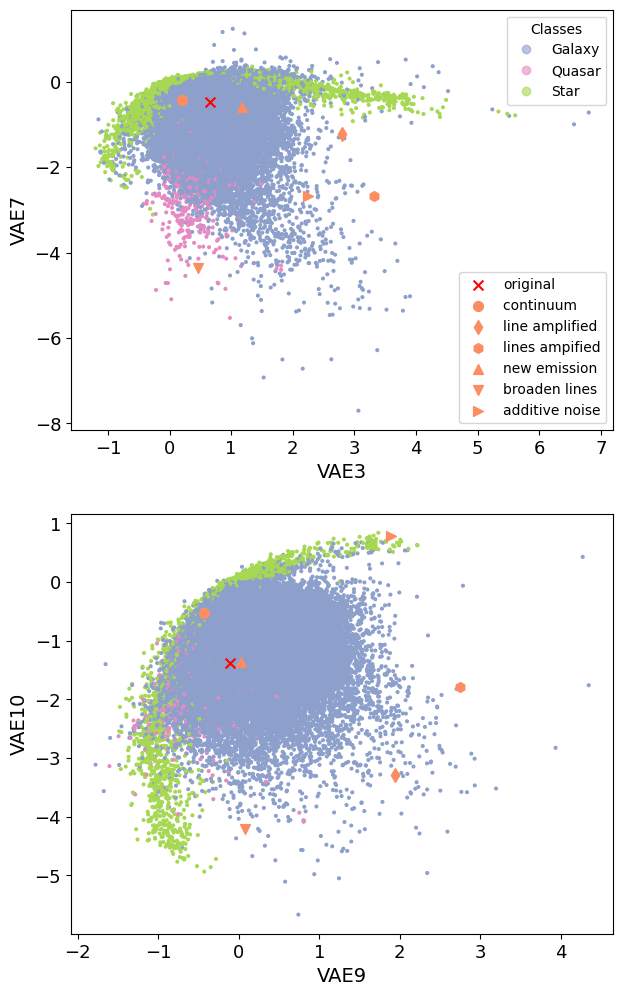}
\caption{Scatter plots of the training spectra in two planes of the latent latent space, components 3 and 7 (top plot) and 9 and 10 (bottom plot). Points are colour-coded by spectral class: galaxies (blue), stars (green), and quasars (pink). The red cross marks the position of a well-reconstructed (unperturbed) galaxy spectrum. The orange markers indicate synthetically perturbed versions of this spectrum.}
\label{fig:artificial_anomalies}
\end{figure}

To probe how the trained VAE represents departures from normal spectral features, we injected controlled synthetic anomalies into a single galaxy spectrum and tracked their locations in latent space. This experiment provides intuition for how different spectral perturbations are encoded and which deviations are considered anomalous by the model.
We selected a galaxy spectrum that is well reconstructed by the VAE ($\textrm{MSE}=0.88$) and lies within a dense region of the latent space. We then generated modified versions of this spectrum to represent a variety of anomalies:

\begin{enumerate}
\item Continuum only: the spectrum was smoothed with a median filter to remove emission and absorption features, retaining only the continuum shape.
\item Amplified single line: the existing \ha\ line was enhanced by a factor of 5.
\item Amplified multiple lines: \ha\ , \oii\ and \hb\ were enhanced by factors of 5, 5, and 8, respectively.
\item New emission lines: narrow emission lines were added at $4200$, $5500$, $7100$, and $7400~\text{\r{A}}$ with amplitudes comparable to the spectrum’s \ha\ line.
\item Broadened lines: \ha\ and \hb\ lines were synthetically broadened.
\item Added noise: Gaussian noise was added with an amplitude comparable to half the total spectral range.
\end{enumerate}

Each perturbed spectrum was processed through the same normalisation pipeline as the training data and encoded by the trained VAE. Figure~\ref{fig:artificial_anomalies} shows their latent projections overlaid on the training set distribution. We show two planes of the latent space; latent components 3 and 7 in the top plot and 9 and 10 in the bottom plot. The red cross indicates the position of the original spectrum and the orange markers indicate the positions of the perturbed spectra. We colour-code the projected training data according to their spectral classification. For an in-depth analysis of the VAE's latent space, including all pairwise 2D projections of the 10D latent space see Section~\ref{sec:inter_latent_space}.

The continuum-only spectrum shifted modestly in latent space due to the loss of emission-line information, but remained within a populated region since many real galaxies display smooth continua. Its reconstruction error decreased slightly ($\textrm{MSE}=0.72$), indicating that the VAE captures continuum shapes accurately; such spectra would therefore not be flagged as anomalous.

In contrast, amplifying emission lines displaced the spectrum into increasingly underdense regions of the latent space. Strengthening only \ha\ produced a noticeable offset, while enhancing multiple lines caused an even larger shift. This demonstrates the VAE’s sensitivity to atypical line strengths and ratios: such perturbations move spectra away from the learned correlations that define the normal spectral manifold.

Adding new narrow lines caused only a small latent displacement, but resulted in a very large reconstruction error ($\textrm{MSE}=30.3$). As the encoder has never encountered similar line features, it projects the spectrum near familiar regions of latent space, while the decoder fails to reproduce the unseen lines leading to high reconstruction error. This highlights the importance of using complementary metrics for robust anomaly detection.

The broadened-line spectrum moved toward a region of latent space occupied primarily by quasars. This is more evident in the top plot in Figure~{\ref{fig:artificial_anomalies}} where the separation of the quasar class is more prominent. Despite being trained without labels, the VAE organises spectra according to shared characteristics, in this case, the presence of broad emission lines, demonstrating that physically meaningful structure emerges naturally in the latent representation.

Finally, adding Gaussian noise pushed the spectrum into a sparsely populated region of latent space, consistent with the model interpreting noisy or corrupted inputs as atypical.

With the exception of the continuum-only case, all perturbed spectra yield large reconstruction errors ($\textrm{MSE} > 10$), demonstrating the VAE’s sensitivity to unseen spectral features and its strong capacity for anomaly detection.
These experiments show that the VAE captures the main modes of spectral variability and organises spectra in a physically meaningful way, with latent dimensions reflecting properties such as emission-line strength and width. The model responds to unusual or unseen features through shifts into underdense latent regions or high reconstruction errors, demonstrating the complementarity of these two anomaly metrics. However, its tendency to map unfamiliar inputs onto nearby normal regions highlights a limitation: anomalies may remain undetected if relying only on metrics that measure isolation in the latent space. Overall, these tests confirm the VAE’s capacity to distinguish both physical and instrumental anomalies while offering interpretable insights into spectral features, emphasizing the value of combining multiple detection criteria for robust anomaly identification.

\subsection{Astronomaly}\label{sec:astronomaly}
While our VAE approach to anomaly detection has drastically reduced the number of spectra an astronomer would need to go through to identify anomalies, a common issue remains. As we have seen, the anomalies fall into many categories. An astronomer who is interested in spectra with transient event signatures may have little interest in outliers that were flagged due to artefacts or wrongly assigned redshifts, whereas an astronomer in charge of data quality will be interested in the latter. The relevance of the outliers is inherently subjective and is based on the scientific goal. The solution we adopted in this study is Astronomaly \citep{Lochner2021astronomaly}. Astronomaly offers a solution to the outlier relevance problem by using Active Learning to offer personalised curation of outliers to the user.

There are three main steps to the Astronomaly framework: feature extraction, anomaly detection, and Active Learning. Feature extraction refers to the dimensionality reduction of high dimensional data to a few components that contain the most critical information. In our case, this is the latent space provided by our VAE and hence the feature extraction in Astronomaly's framework is skipped. Next, Astronomaly applies the LOF algorithm to the latent space provided, similarly to the methodology we demonstrated earlier in this study. The spectra are ranked according to lowest NOF and presented in this sorted order to the user in Astronomaly's interface. We adapted the interface to tailor it to our use case, for optimal plotting of spectra. Astronomaly also allows the user to observe where spectra reside in the latent space by using t-distributed Stochastic Neighbor Embedding (t-SNE, \cite{vanDerMaaten2008}) to project the $10$D latent space to $2$D. 
This second step of Astronomaly which involves calculating the anomaly score using the LOF algorithm could potentially be replaced with the reconstruction error, or more appropriately reconstruction probability method. Alternatively, the two approaches could also be combined taking into account both the position on the latent space and reconstruction probability.

In the third step, the expert can visually inspect the sorted spectra (from highest to lowest anomaly score) and provide a label ($0$-$5$) that reflects how relevant, or not, the specific outlier spectrum is to the user. The novelty Astronomaly introduces to our analysis is through this last step. Active Learning is used to adjust the original anomaly scores provided by LOF according to the relevance score provided by the user labels, learning in this way which anomalies the user is interested in. To predict the relevance score for the remainder of the dataset, Astronomaly uses a random forest regression algorithm \citep{Breiman2001RF}. The original anomaly score of a data point is updated based on the relevance score predictions and its distance in latent space to the nearest human-labelled neighbour. For regions with few to no human labels, the score reverts to the original anomaly score provided by LOF which means that the user gets the opportunity to see and label the object, so new anomalies are less likely to be missed. In regions where confidence is high, the predicted relevance score dominates. When the Active Learning is trained on the human labels, the outliers are reordered according to the new relevance anomaly scores prioritising in this way outliers that are more relevant to the user. The process can be iterated as more human labels are added to provide improved outlier recommendations. 

Astronomaly removes the need to define an explicit anomaly threshold on the raw scores. Instead, it employs active learning to surface the most relevant spectra for the user. This reduces the number of false positives and increases the likelihood that the anomalies presented to the user are scientifically meaningful, thereby improving efficiency compared with fixed-threshold methods.

We applied Astronomaly to the VAE encoded DESI spectra. To qualitatively examine the effect of the Active Learning process, we assigned a high relevance score label to spectra with a discontinuity due to miscalibration of the blue and red channels and to extreme emission lines (similarly to Figure~\ref{fig:mse_outliers} (a) and Figure~\ref{fig:nof_outliers} (c)). Providing labels to only the first $50$ spectra shown and training the Active Learning step on these, we observed that the updated sorted list of outliers is now mostly composed of these two classes of spectra.

\subsection{Interpretation of the Latent Space}
\label{sec:inter_latent_space}
%%%%%%%% CORNER PLOT
\begin{figure*}
    \centering
    \includegraphics[width=1\textwidth]{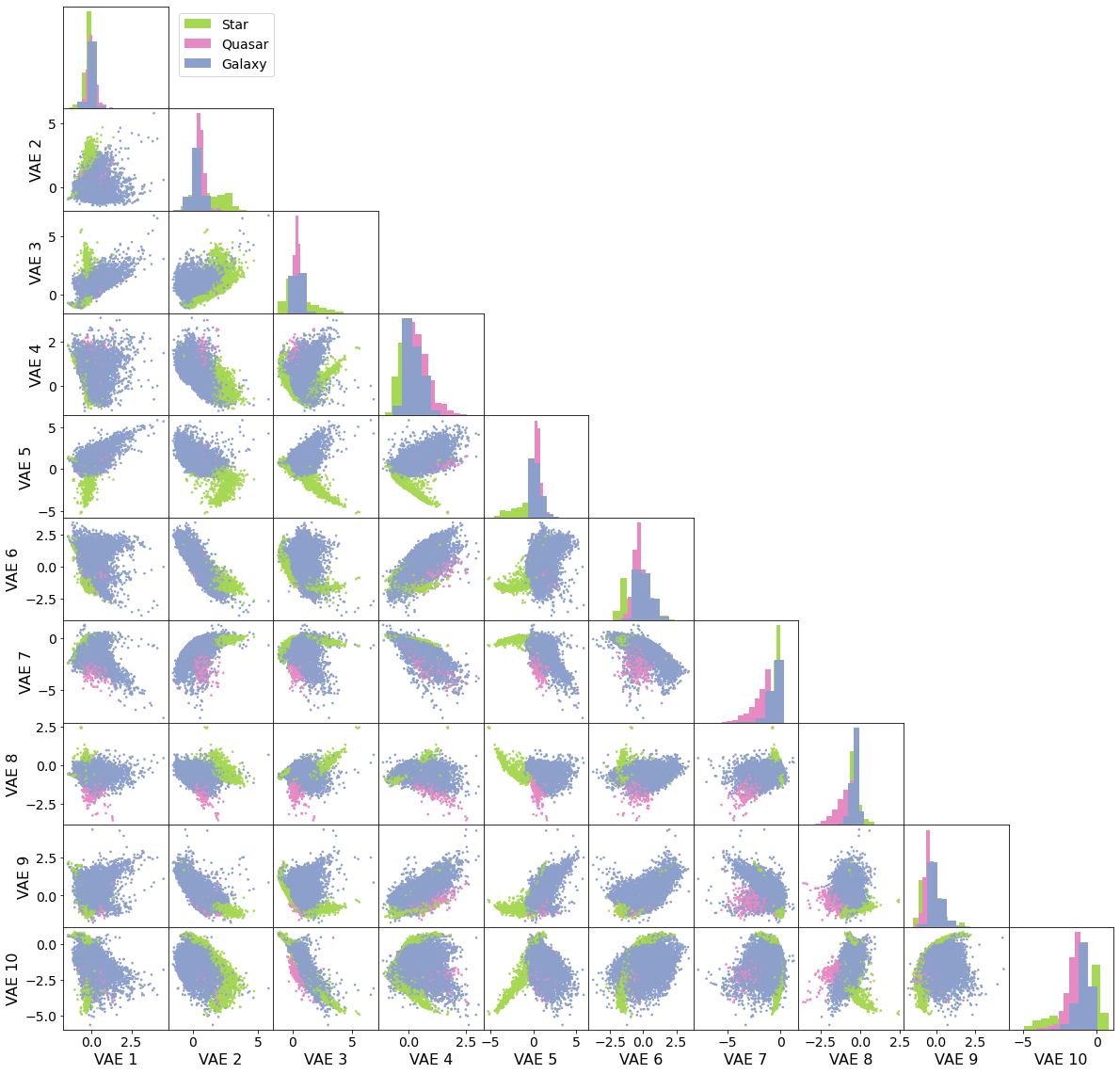}
    \caption{Corner plot of the training spectra embedded in the $10$ dimensional VAE latent space. The spectra are colour-coded according to their spectral classification.} 
    \label{fig:corner_plot}
\end{figure*}
In addition to anomaly detection, one of the great advantages of VAEs is that they enable us to learn more about the bulk of the distribution through examination of the latent space.
In Figure~\ref{fig:corner_plot} we plot the latent representation of the training spectra across all $10$ VAE latent dimensions using a corner plot. While these pairwise $2$D projections of the $10$D latent space do not capture the distribution of spectra in its totality, the scatter plots are still highly informative and useful when trying to interpret the latent space. The data are colour-coded according to their spectral classification. We observe that the VAE successfully separates the three classes without ever having seen the labels. The labels are strictly used for colour-coding plots only. The separation is particularly good for stars, evident in the fifth latent dimension for example. The scatter plot of VAE $5$ and VAE $10$ shows two distinct groups, the galaxies in blue and the stars in the green extending arm. Quasars are also separated, shown more clearly in the seventh and eighth dimensions. The scatter plot of VAE $5$ and VAE $8$ clearly shows the quasars represented in pink separating from the galaxies and stars.

Beyond examining the class separation in latent space, it is insightful to also understand how sub-classes of spectral classes are distributed in the latent space. As we have not included any information on sub-classes, the way that we approach this is by examining the pairwise $2$D projections of the $10$D latent space focusing on a single class each time and extracting spectra, along a specified track in the latent space. 
%Observing the actual spectra along a latent space track we can obtain an understanding of the distribution of spectra in the latent space. 

\subsubsection{Galaxy Tracks}
\begin{figure}%[t!]
    \centering
    \includegraphics[width=0.45\textwidth]{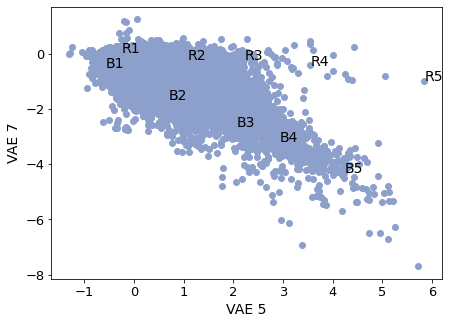}
    \caption{Scatter plot of the fifth and seventh VAE dimensions of the \textbf{galaxy} spectra only. The labels draw out two tracks, B and R, which correspond to bluer and redder continua in spectra.} 
    \label{fig:vae5_vae7}
\end{figure}

\begin{figure}%[t!]
    \centering
    \includegraphics[width=0.48\textwidth]{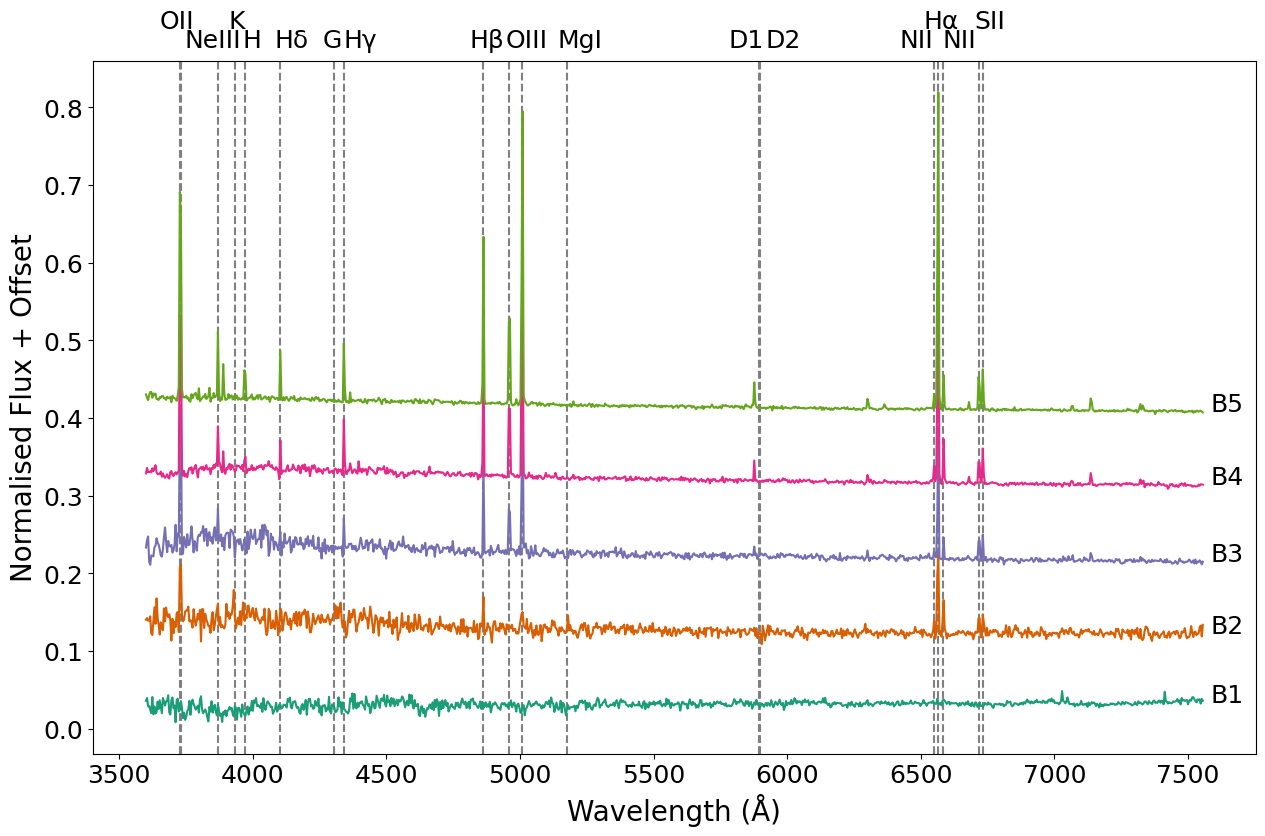}
    %\caption{} 
    %\label{fig:gal_trackA}
%\end{figure}

%\begin{figure}%[h!]
    \centering
    \includegraphics[width=0.48\textwidth]{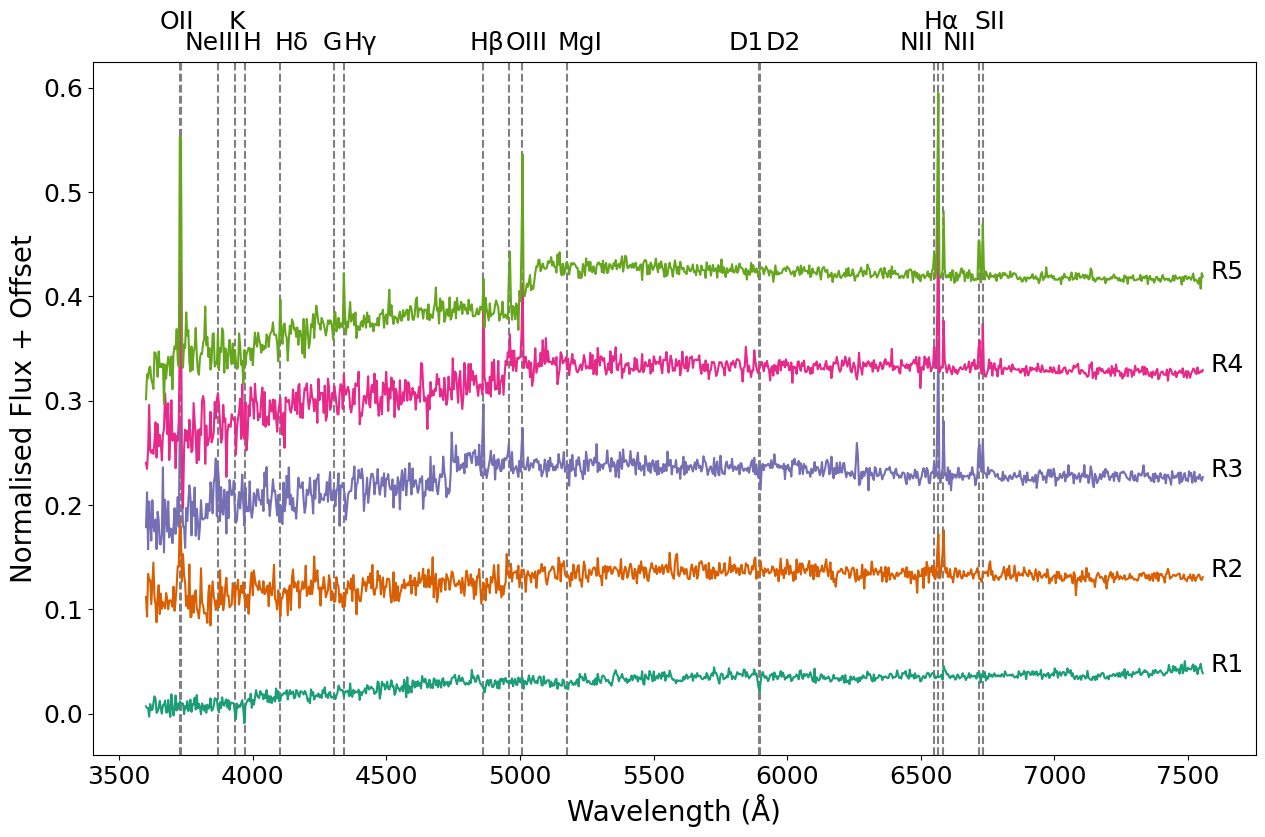}
    \caption{Galaxy spectra along the B (top) and R (bottom) tracks drawn in the VAE latent space shown in Figure~\ref{fig:vae5_vae7}. The normalised spectra are plotted with an arbitrary offset for visibility. The R track shows redder spectra whereas the B track shows bluer spectra.} 
    \label{fig:gal_trackA+B}
\end{figure}

Figure~\ref{fig:vae5_vae7} shows the $2$D scatter plot of VAE $5$ and VAE $7$ for the galaxy spectra only. From this plot we can observe a main tail extending to the lower right side of the plot (indicated by following the ``B" labels) and a weaker extension to the upper right side of the plot (indicated by the following the ``R" labels). The tracks have been labelled R and B because the constituent spectra exhibit redder and bluer continua correspondingly. We extract an indicative galaxy spectrum from each point along the paths B$1$-B$5$ and R$1$-R$5$ and plot these in Figure~\ref{fig:gal_trackA+B} (top and bottom correspondingly). 
%The top plot in Figure~\ref{fig:gal_trackA+B} shows the galaxy spectra corresponding to B$1$-B$5$ plotted with an offset to avoid overcrowding, and the bottom plot shows the galaxy spectra corresponding to the points R$1$-R$5$ in the latent space. 

%%%% Tracks GAL

We make sure that the galaxy spectrum corresponding to a point in latent space is representative of the majority of the spectra surrounding that point by inspecting several spectra that lie close to it. Another approach to ensure this, would have been to plot the stacked spectra of the galaxy spectra that fall within a small radius from the specified point in the latent space. Stacking spectra helps to better study the average properties of the galaxies involved and reveal fainter spectral signatures.

The R$1$ spectrum has absorption lines at \caii\ H and K and \nai\ (D-lines) and a red continuum with a $4000~\text{\r{A}}$ break indicating an old metal-rich stellar population. All spectra in the R-track show a decrease in flux in the blue part of the continuum. R$2$ shows emission lines in \oii, \ha\ and \nii. As we follow the R track these emission lines are strengthened and joined by \sii, \oiii\ and the rest of the Balmer series indicating star formation. The \ha\ emission line increases in strength as we follow the R track. 
%%%%%%%%%%%
%%%%%%%%%%%%%
The R$5$ spectrum shows a strong discontinuity around $5000~\text{\r{A}}$ which is in the wavelength band where the data from the blue and red cameras of the DESI spectrograph are coadded. After inspecting the individual camera data we observed that the continuum flux around $5000~\text{\r{A}}$ recorded by blue camera was shifted down relative to that of the red camera and therefore this discontinuity is likely an artefact arising from a miscalibration between the two channels. 
Interestingly, the R$5$ spectrum is isolated in the latent space indicating that the VAE has picked up on this atypical feature. 
Inspecting the spectrum closest to R$5$ in the latent space we observed that it also exhibits a, slightly weaker, discontinuity at $5000~\text{\r{A}}$. 

The B$1$ galaxy spectrum similarly to the R$1$ spectrum, does not have any strong emission features. In contrast to the R$1$ spectrum, its continuum is bluer and it does not exhibit the $4000~\text{\r{A}}$ break indicating a younger stellar population. B$2$ is also a blue spectrum and has emission lines in \oii, \hb, \ha, \nii\ and \sii. Following the B track, these lines increase in strength and are joined by the rest of the Balmer emission lines, \oiii\ and a weak \hei\ emission line. The blue part of the spectra shows several emission lines. B$3$-$5$ exhibit strong \ha\ emission with \ha\ $\gg$ \nii\ suggesting low-metal starburst galaxies \citep{Almeida_2012}. Again we can see that the \ha\ emission increases in strength as we follow the B track.

\subsubsection{Quasar Tracks}
%%%% Tracks QSO
\begin{figure}%[b!]
    \centering
    \includegraphics[width=0.45\textwidth]{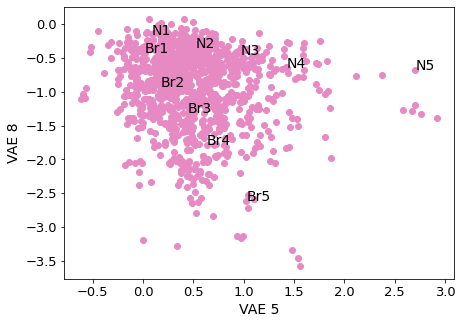}
    \caption{Scatter plot of the fifth and eighth VAE dimension of the \textbf{quasar} spectra only. The labels draw out two tracks, Br and N, which correspond to broad and narrow spectra.} 
    \label{fig:vae5_vae8}
\end{figure}
%%%%%%%%%%%%%%%

\begin{figure}%[t!]
    \centering
    \includegraphics[width=0.48\textwidth]{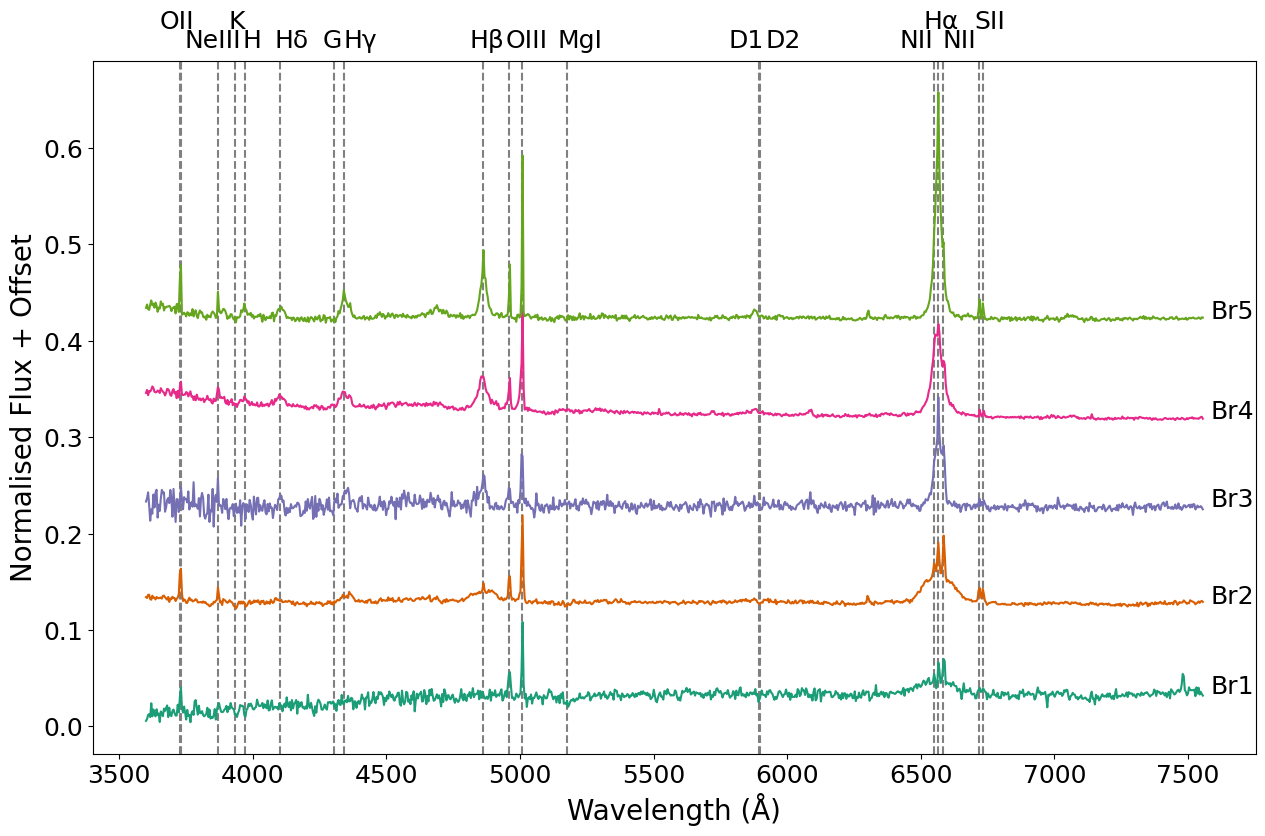}
    %\caption{} 
    %\label{fig:qso_trackA}
%\end{figure}

%\begin{figure}%[h!]
    \centering
    \includegraphics[width=0.48\textwidth]{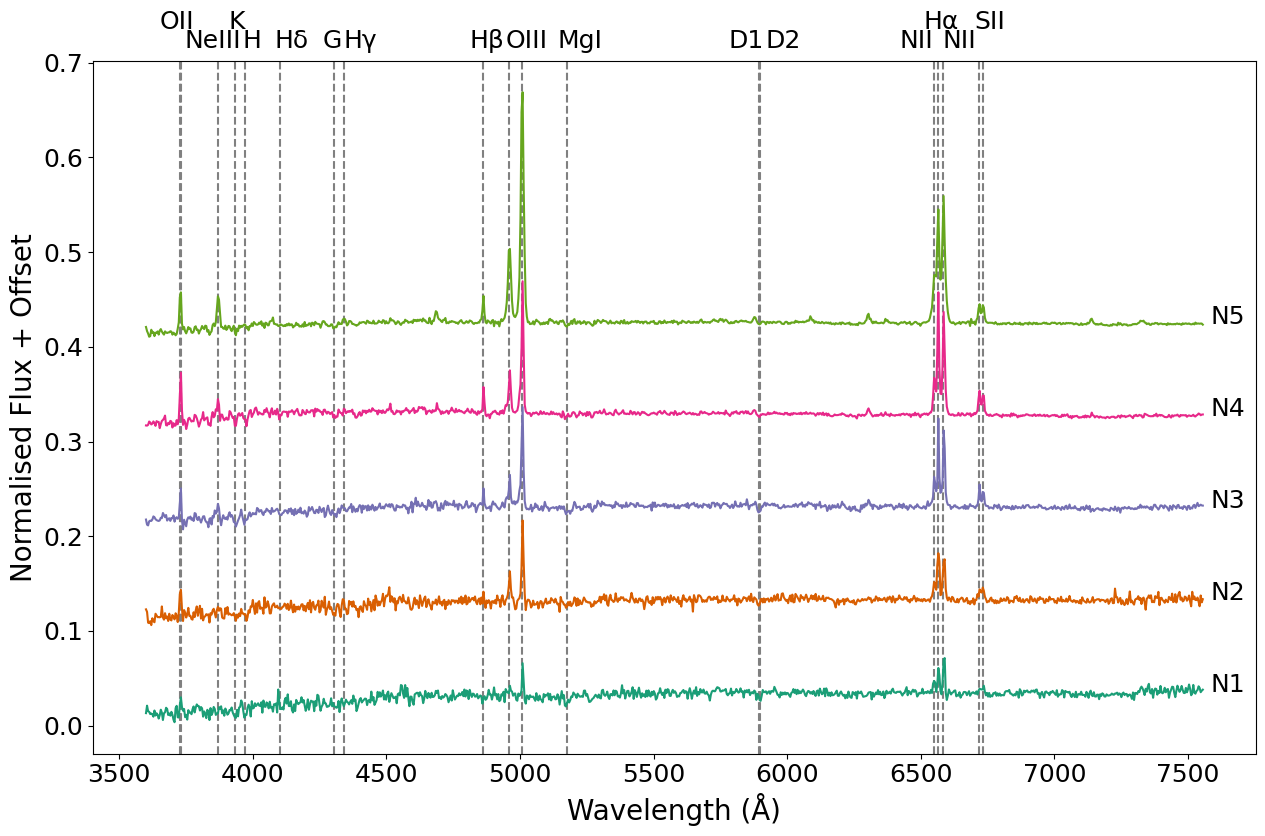}
    \caption{AGN spectra along the Br (top) and N (bottom) tracks drawn in the VAE latent space shown in Figure~\ref{fig:vae5_vae8}. The normalised spectra are plotted with an arbitrary offset for visibility. The Br track shows broad spectra whereas the N track shows narrower spectra.} 
    \label{fig:qso_trackAB}
\end{figure}

We repeat the analysis for quasar spectra. In Figure~\ref{fig:vae5_vae8} we plot the $2$D scatter of VAE $6$ and VAE $8$ for the quasar spectra only. We draw out two tracks, labelled Br (Broad) and N (Narrow) corresponding to two extensions from the main collection of scatter points.
The spectra extracted along the two tracks are plotted in Figure~\ref{fig:qso_trackAB}, where the top plot shows spectra from the Br track and the bottom plot shows spectra from the N track.

%%%%%%%%%%

The Br$1$ spectrum shows emission lines in \oii, \oiii, \ha\ and \nii. We can observe that there is broadening in the spectrum around the \ha\ emission line. As we follow the Br track, the \ha\ and \oiii\ emission lines, in particular, increase in strength and decrease in broadening. Broad emission lines for the rest of the Balmer series are also visible, especially in Br$4$ and $5$. We also observe that the spectra exhibit a blue continuum. As the spectra in the Br track have both broad and narrow spectral lines the most likely source is Seyfert 1 galaxies. 

The N track on the other hand, has only narrow emission lines.
The N$1$ spectrum has emission lines in \oiii, \ha\ and \nii. Subsequent spectra in the N track show increasing strength in these lines as well as emission lines in \oii, \neiiilam, \hb\ and \sii.
The \oiii\ emission is particularly strong with \hb\ being much weaker. We also notice that for these galaxies \ha\ is of similar strength to \nii. These characteristics suggest that these spectra correspond to Seyfert 2 galaxies. 

\subsubsection{Star Tracks}
\begin{figure}%[b!]
    \centering
    \includegraphics[width=0.44\textwidth]{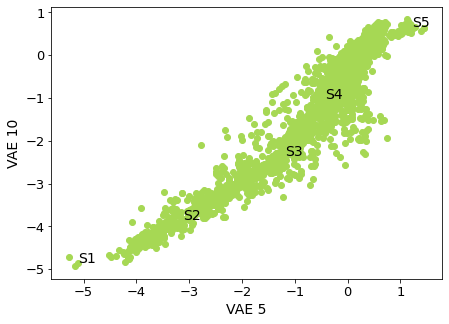}
    \caption{Scatter plot of the fifth and tenth VAE dimension of the \textbf{star} spectra only. The labels draw out a single track labelled S.} 
    \label{fig:vae5_vae10}
\end{figure}

Finally we examine the $2$D scatter plot of VAE $5$ and VAE $10$ for the star spectra only, shown in Figure~\ref{fig:vae5_vae10}. The star spectra projected in this $2$D latent space are arranged in an elongated structure. Here we examine a single track, labelled S, following the distribution of star spectra from the bottom left to the top right of the plot.

%%%% Tracks Stars

%%%%%%%%
\begin{figure}%[h!]
    \centering
    \includegraphics[width=0.48\textwidth]{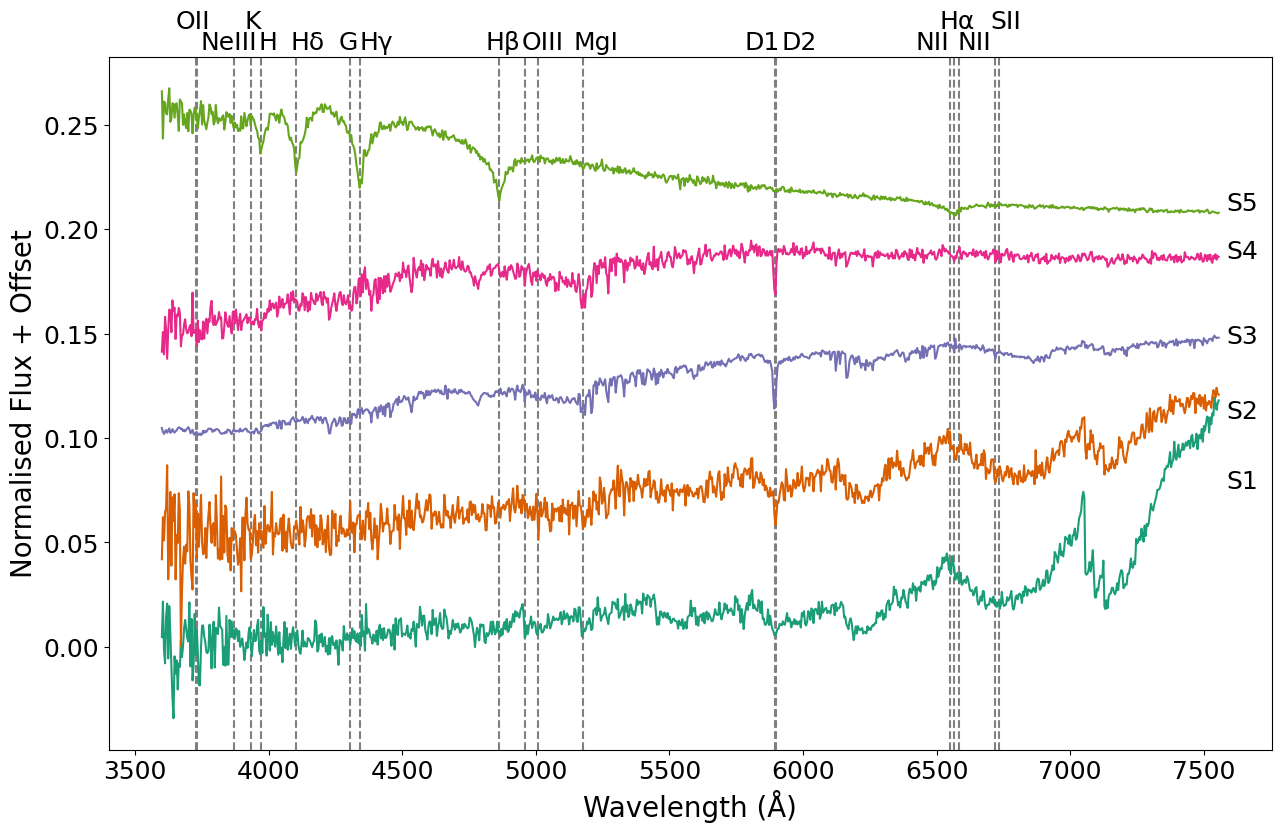}
    \caption{Star spectra along the S track drawn in the VAE latent space shown in Figure~\ref{fig:vae5_vae10}. The normalised spectra are plotted with an arbitrary offset for visibility.} 
    \label{fig:star_track}
\end{figure}

The spectra are plotted in Figure~\ref{fig:star_track}.
The S$1$ and S$2$ spectra are from low temperature stars with a rightward rising profile that peaks in the red/infrared. We can also observe absorption in the \nai\ D-lines and the TiO (Titanium oxide) bands at $\lambda4955, 7150$. These features are typical of M stars. Note how similar the spectrum in Figure~\ref{fig:star_track} in dark green, which corresponds to the S$1$ point in Figure~\ref{fig:vae5_vae10}, is to the M star with high flux in the red part of the spectrum identified in Figure~\ref{fig:nof_outliers} (b). S$3$ and S$4$ show stronger absorption in the \nai\ D-lines and other neutral metals such as \mgi. The continuum again shows a rise in the red part of the spectrum but the gradient is less steep than that of M stars. The spectra at S$3$ and S$4$ are typical of K stars. The latent space distribution of the stellar spectra suggests that the VAE has successful separated K and M stars, with M star spectra found at the bottom half of the elongated structure in the latent space and K stars at the upper part.
Finally we inspect the region found at the very top of the plot that is slightly separated from the main body, represented by S$5$. The spectrum at S$5$ is significantly bluer and exhibits strong absorption lines mainly from the hydrogen atom (Balmer series). The absorption lines are significantly broadened, indicating that this is a white dwarf. 
%The spectral lines of white dwarfs are broadened primarily due to the Stark effect which postulates that the allowed energy levels are shifted when the atom is placed in an external electric field with the shift being dependent upon the strength of the field. This causes the emitted photons to have a range of wavelengths around the wavelength of the level transition in the absence of the electric field.  Electric fields in white dwarfs originate from high temperatures which cause some of the atoms in the atmosphere of the star to become ionised. As white dwarfs have atmospheres that are $10,000$ times denser than main sequence stars, these free electrons and ions are close enough to the neutral emitters for them to experience the presence of the electric field.
While inspecting the surrounding area of S$5$ in the latent space, we also observed stellar spectra similar to the one corresponding to a white dwarf but without broadening of the spectral lines. These are spectra originating from hot, blue stars such as B and A stars. These have been placed by the VAE close to the white dwarf spectra in latent space, as they all exhibit very a blue spectral profile with strong absorption lines from the hydrogen atom. 

\subsubsection{VAE Generated Spectra}

\begin{figure*}%[h!]
    \centering
    \includegraphics[width=0.8\textwidth]{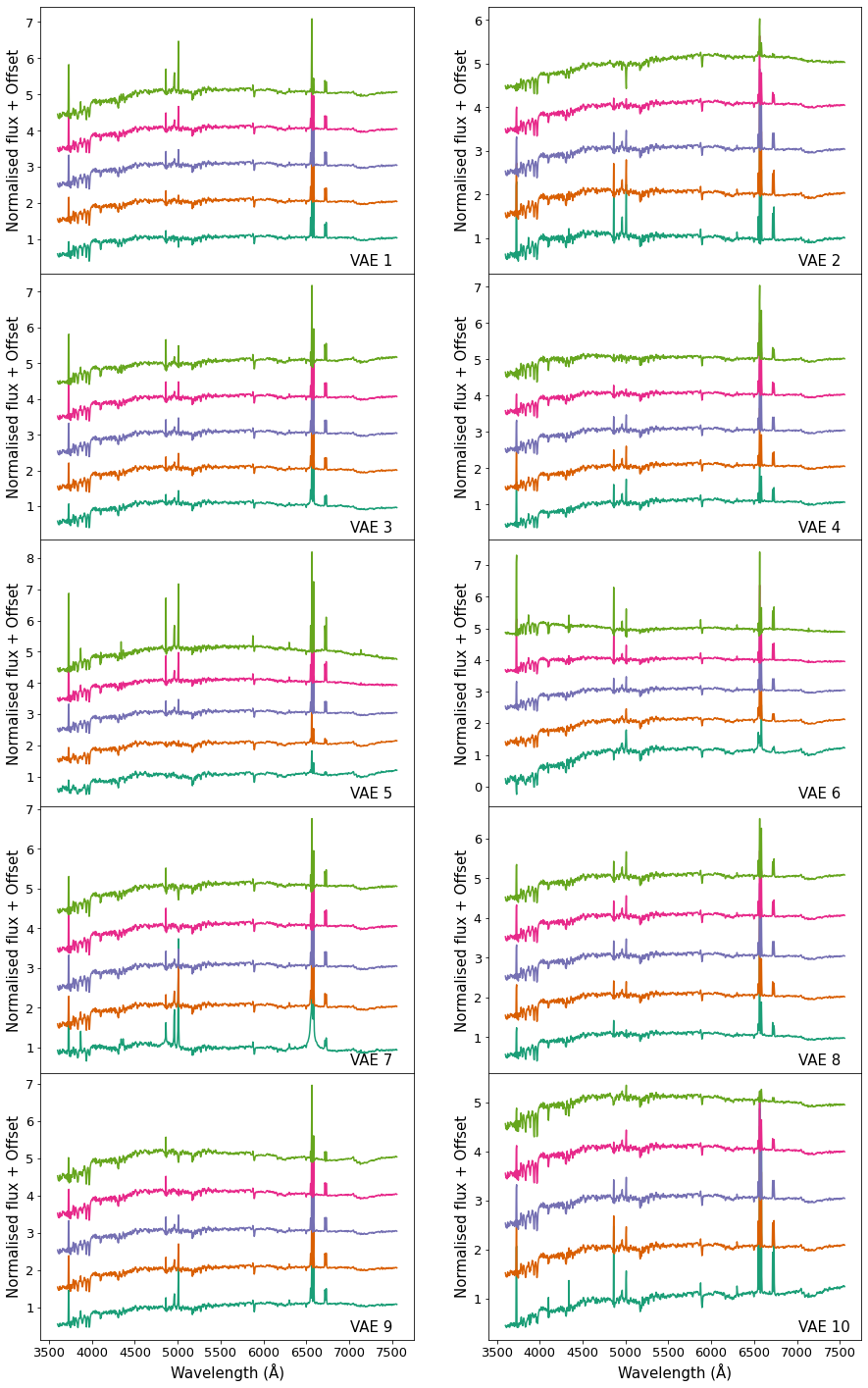}
    \caption{Synthetic spectra generated using the VAE, along each VAE latent dimension (in the increasing direction) while keeping the rest of the VAE dimensions fixed to their means. The middle spectrum in each subplot is common in all and corresponds to the centroid of the VAE latent means. The spectra are plotted with offset for visibility.}
    \label{fig:generated_spectra}
\end{figure*}

In order to better understand the information captured by individual VAE components, and to augment the insights we have already gained by looking at the synthetic anomalies in Section \ref{sec:artifial_anomalies}, we generate spectra, using the trained VAE by sampling from the latent space along individual VAE latent dimensions. The coordinates of the generated spectra in the latent space are fixed to the mean value of each latent dimension, with only one dimension being varied at each time. We repeat this for all VAE latent dimensions. The results are plotted in Figure~\ref{fig:generated_spectra}.
The middle spectrum of each subplot in Figure~\ref{fig:generated_spectra} plotted in purple is common to all and represents a spectrum generated from the centroid of all spectra's latent means.

\begingroup
\setlength{\tabcolsep}{10pt} % Default value: 6pt
\renewcommand{\arraystretch}{1.5} % Default value: 1
\begin{table*}%[b!]%\label{tab:cosmo_param}
\caption{Summary of how the main spectral features change as we move along the specified VAE latent dimension (L. D.), in the increasing direction, as depicted in Figure~\ref{fig:generated_spectra}. The dash indicates that the feature remains approximately constant for that VAE dimension, the $\uparrow$ indicates that the feature increases in strength and the $\downarrow$ indicates that it decreases.}
    \centering
    \begin{tabular}{>{\centering\arraybackslash}m{1.2cm}>{\centering\arraybackslash}m{2.2cm}>{\centering\arraybackslash}m{0.3cm}>{\centering\arraybackslash}m{0.5cm}>{\centering\arraybackslash}m{0.3cm}>{\centering\arraybackslash}m{0.5cm}>{\centering\arraybackslash}m{0.5cm}>{\centering\arraybackslash}m{2cm}}
         \hline\hline
         VAE L. D.& Continuum & \ha & \nii\ & \hb & \oiii & \oii & Broadening \\
         \hline
         1  & - & $\uparrow$   & $\downarrow$ & $\uparrow$   & $\uparrow$   & $\uparrow$   & No \\
         2  & - & $\downarrow$ & $\downarrow$ & $\downarrow$ & $\downarrow$ & $\downarrow$  & No\\
         3  & - & $\uparrow$   & -            & $\uparrow$   & -            & $\uparrow$    & No\\
         4  & - & $\uparrow$   & $ \uparrow$  & $\downarrow$ & $\downarrow$ & $\downarrow$  & No\\
         5  & red $\downarrow$ & $\uparrow$ & $\uparrow$ & $\uparrow$ & $\uparrow$ & $\uparrow$ & No\\
         6  & blue $\uparrow$, red $\downarrow$, $4000~\text{\r{A}}$ weakens & $\uparrow$ & $\downarrow$ & $\uparrow$ & - & $\uparrow$ & Yes \\
         7  & blue $\downarrow$ & - & - & - & $\downarrow$ & -  & Yes \\
         8  & - & $\downarrow$ & - & - & $\uparrow$ & - & No\\
         9  & - & $\uparrow$ & $\downarrow$ & - & $\downarrow$ & $\downarrow$ & No\\
         10 & red $\downarrow$ & $\downarrow$ & $\downarrow$ & $\downarrow$ & $\downarrow$ & $\downarrow$ & No\\
         \hline\hline
    \end{tabular}
    \label{tab:VAE_components_description}
\end{table*}
\endgroup

%%%%%%%%%% Generated spectra along VAE dimensions
%%%%%%%%%%%
To assist in the interpretation of Figure~\ref{fig:generated_spectra}, in Table~\ref{tab:VAE_components_description} we summarise the changes to the main spectral features that occur, as we move along each latent dimension (in the increasing direction). 
The continuum is observed to remain unchanged in shape and amplitude in most latent dimensions. The red region of the spectrum is observed to decrease in VAE $5$, $6$ and $10$ with $6$ also exhibiting a weakening in the $4000~\text{\r{A}}$ break along with an increase in the blue region of the spectrum. In contrast, VAE $7$ shows a decrease in the blue part of the spectrum which we observed earlier in Figure~\ref{fig:vae5_vae7} where redder spectra had higher values in this dimension. These dimensions capture information about the age of the dominant stellar population. VAE $6$ and $7$ are also able to describe broadened emission lines as we can observe larger widths in the Balmer series lines. 
VAE $2$ and $10$ both show decrease in all emission lines leading to a spectrum with mostly absorption lines. VAE $4$ also shows decrease in all emission lines, except \ha\ and \nii\ which increase in strength. Conversely, VAE $5$ shows an increase in all emission lines. Both VAE $5$ and $10$ show a strong TiO band at $\lambda7150$ which is expected as low values of both VAE $5$ and $10$ are used to describe M stars. We tentatively observe also that decreasing VAE $5$,  seems to decrease star formation and increase stellar age.

Studying the VAE latent dimensions individually allows us to interpret these and relate them to the physical attributes of the spectra. However, we should be aware that these interpretations change when multiple dimensions are changed from their mean at the same time. Collectively, the VAE dimensions, give the VAE the flexibility to reconstruct complex spectra by making use of non-linear combinations. In addition, any physical trends suggested by this work are intended to be indicative and would need to be confirmed by traditional techniques such as stellar population synthesis fitting.

\section{Discussion and Conclusions}\label{sec:discussion}

In this paper we have:

\begin{itemize}
    \item used a VAE to detect anomalous DESI spectra, finding them as either outliers in the VAE latent space or those which are reconstructed poorly compared with their original representation;
    \item used Astronomaly to curate these sets of outliers further making visual inspection more effective; and
    \item explored the VAE latent space with both synthetic anomalies and spectral tracks, showing how it can be used for unsupervised classification, anomaly detection and that it can encode physical characteristics of the dataset.
\end{itemize}

The increasing complexity and volume of astronomical spectra, requires the use of data intensive techniques. In this work, we have shown that a VAE is capable of effectively compressing the information found in spectra by a factor of $100$, whilst retaining enough information to accurately reconstruct the different features present in spectra as shown in Section~\ref{sec:recon_acc}. PCA has also been shown to effectively compress spectra \citep{RonenLahav1999pca, Madgwick2003, Portillo2020vae}, although, \cite{Yip_2004b_quasarclass} demonstrated its limitation when non-linear features, such as broad emission lines, are present.
%The VAE hyperparameters used in this study are not the optimal hyperparameters as they have been chosen from a limited random search.
Optimising the VAE hyperparameters used in this study by way of a random or grid search should lead to better performance and more accurate reconstructions. 
%In this study we have shown that VAEs are capable of effectively compressing the information found in spectra by a factor of $100$. 
%In Section~\ref{sec:recon_acc}, have shown that a VAE with $10$ latent dimensions achieves good reconstructions of the different features present in spectra.
%The fully-connected layers could also be replaced with convolutional layers similarly to the approach by \cite{Teimoorinia_2022}.
%\cite{Teimoorinia_2022} use a Deep Embedded Self-organising Map (DESOM), comprised of an Autoencoder (AE) with convolutional layers with the aim to map integral field unit (IFU) spectra from the Manga survey onto a map according to similarity. 
%and with a latent dimension of $256$ (original dimension of spectra is $4544$).  The MSE is used to evaluate the reconstuction quality The reduced feature space is then passed to a self-organising map (SOM) that further clusters similar spectra. The two algorithms are trained simultaneously. The output of DESOM is a ``fingerprint" for every object corresponding to the projection in the SOM the spectra within a single galaxy fall onto our DESOM map in a unique pattern. This revealed the presence of distinct stellar populations, indicated star formation histories and related these to galaxy morphology.
%concolutional VAE - aid the extraction of correlated features from the spectra - tEIMOORINIA 

The VAE successfully managed to group together spectra belonging to the same class and separate the different classes in latent space, without having ever seen the corresponding labels. The VAE encoder produces a latent space that is interpretable. 
By injecting synthetic spectra into the dataset, we demonstrated the VAE’s capacity to distinguish both physical and instrumental anomalies while offering interpretable insights into spectral features. We took this further by using spectral tracks to examine
 the latent space. In so doing, we found the VAE was able to distinguish between bluer and redder continua and broad and narrow emission lines. Traversing different tracks, we observed that the spectra are distributed in such a way that there is a gradual evolution in the strength of emission and absorption lines. The changes between adjacent points in the latent space are not abrupt, providing evidence that the VAE latent space created is indeed regularised. We found tracks that correspond to increasing star formation and increase in broad emission lines along the Balmer series. The VAE also successfully grouped different sub-groups of stellar spectra. 
Using star sub-classes and line-ratio classifications such as BPT (\citealt{Baldwin_1981_BPT}) or its updates to obtain galaxy sub-classes will further aid in the interpretation of the latent space. Unlike, line-ratio diagnostics, the VAE also captures the information present in the continuum.

The generative properties of the VAE allow us to probe different points in the latent space and interpret the information encoded in each dimension. When generating from the latent space care should be taken to primarily sample from places in the latent space that are occupied by training examples. Sampling far away from the encoded training spectra, forces the VAE to extrapolate often resulting in non-meaningful outputs.

A non-linear dimensionality reduction algorithm such as t-SNE or UMAP \citep[Uniform Manifold Approximation and Projection,][]{McInnes2018UMAP} which can preserve the global structure of the $10$D latent space, could be used to further compress the encoded information to $2$D. Such methods facilitate visualisation of high dimensional data and allow for a more efficient exploration of the latent space than inspecting a corner plot of all latent dimensions. \cite{Pat2022DESISDSS} applied a Probabilistic Autoencoder \citep[PAE,][]{boehm2022pae, böhm2023fastefficientidentificationanomalous} to reduce the dimensionality of SDSS spectra from $1000$ to $10$ dimensions. UMAP was then applied to the compressed data to further reduce the dimensionality to $2$D which allowed for an efficient interpretation of the projected latent space and identification of patterns that translate to the physical properties of spectra. Performing this two stage compression yielded improved class separation compared to using only UMAP.

We have shown that VAEs are capable of detecting anomalous spectra in the dataset using the weighted MSE between the original and reconstructed spectrum. 
An improvement to the weighted MSE approach would be to use the weighted reconstruction probability instead. As the VAE is a stochastic generative model that outputs probabilities, using the reconstruction probability as a proxy for anomaly detection is more principled and objective. The reconstruction probability is the Monte Carlo estimate of the reconstruction loss in the VAE's loss function \citep{AnCho2015VaeAD}. It takes into account the variability of the distributions from the sampling procedure.

Spectra were also scored according to their position on the latent space using the LOF algorithm. While some spectra were identified by both methods as outliers, LOF identified on average lower S/N outlier spectra. This suggests that a spectrum's position in the latent space is affected by the noise present.
Similarly, bad pixels in spectra will influence the spectrum's position in the latent space, even if their weights are set to zero which prevents them from affecting the reconstruction probability calculation in the loss function. This highlights the importance of treating known bad pixels during the preprocessing by removing them and infilling the missing values using imputation methods, as done in this study. Failing to do so would result in these spectra being identified as outliers due to artefacts we are already aware of. A more comprehensive comparison of the two methods for identifying outliers (MSE and LOF) would be insightful for understanding any trends of outliers identified by each method and their sensitivity to S/N.

In addition to identifying outlier spectra related to rare objects or events, the VAE is also capable of identifying spectra with artefacts or with a wrongly assigned redshift. Both outlier methods employed in this study identified such spectra. \cite{Lan2022DESISV} show that \texttt{redrock} has a high redshift recovery rate from a sample of robust visually inspected redshifts from main survey spectra, with failures being $< 2\%$. This shows that redshift failures are rare and VAEs can facilitate the visual inspection process which is crucial for identifying unexpected problematic features affecting the quality of the data. 
A spectrum that is flagged as an outlier due to a \texttt{redrock} redshift failure or CCD issues which were not identified and properly masked, can help inform of the shortcomings of the DESI spectroscopic pipeline. Based on this, improvements can be made so as to catch any such cases in the future, leading to cleaner data with the majority of outliers being due to physical spectral features. 

The two methods identified atypical spectra in the dataset, including spectra with a wrong assigned redshift, bad pixels, stars, AGNs, miscalibration of the red and blue channels and others.
Whilst we have identified the broad categories corresponding to the top $100$ identified outliers from both methods, domain expertise is required to visually inspect these and determine the significance of the outliers, particularly whether these contain novel events, objects or phenomena that are unknown to scientists. 
To further curate the list of outliers that an expert would need to visually inspect, we made use of Astronomaly which employs Active Learning to provide personalised outlier recommendations based on the user's feedback on which outliers are relevant for a given science question.

Finally, we would observe that VAEs are just one of a family of dimensionality-reduction techniques which could be employed to identify anomalous spectra. In upcoming work, we hope to compare and contrast some of these techniques, in the process proposing enhancements to the standard approaches; and ultimately applying them to find anomalies at scale within the DESI dataset. We note that similar techniques could be applied to other spectroscopic surveys (e.g. Euclid, Roman, Prime Focus Spectrograph, Wide Field Spectroscopic Telescope) and indeed to domains other than astronomy.

\section{Data Availability}
The data underlying the analyses in this paper were drawn from the Early Data Release of the Dark Spectroscopic Instrument (DESI EDR, \citealt{https://doi.org/10.5281/zenodo.7964161}). The full dataset is available at https://data.desi.lbl.gov/doc/releases/edr.

%%%%%%%%%%%%%%%%%%%% ACKNOWLEDGEMENTS %%%%%%%%%%%%%%%%%%
% Using the standard acknowledgements for DESI to be found at
% https://desi.lbl.gov/trac/wiki/GuidetoPub/DESIAcknowledgements

\section*{Acknowledgements}
CN and RPN have been supported by the STFC UCL Centre for Doctoral Training in Data Intensive Science. OL acknowledges STFC Consolidated Grant ST/R000476/1 and visits to All Souls College and the Physics Department, University of Oxford.

The material in this paper is based upon work supported by the U.S. Department of Energy (DOE), Office of Science, Office of High-Energy Physics, under Contract No. DE–AC02–05CH11231, and by the National Energy Research Scientific Computing Center, a DOE Office of Science User Facility under the same contract. Additional support for DESI was provided by the U.S. National Science Foundation (NSF), Division of Astronomical Sciences under Contract No. AST-0950945 to the NSF’s National Optical-Infrared Astronomy Research Laboratory; the Science and Technology Facilities Council of the United Kingdom; the Gordon and Betty Moore Foundation; the Heising-Simons Foundation; the French Alternative Energies and Atomic Energy Commission (CEA); the National Council of Humanities, Science and Technology of Mexico (CONAHCYT); the Ministry of Science, Innovation and Universities of Spain (MICIU/AEI/10.13039/501100011033), and by the DESI Member Institutions: \url{https://www.desi.lbl.gov/collaborating-institutions}. Any opinions, findings, and conclusions or recommendations expressed in this material are those of the author(s) and do not necessarily reflect the views of the U. S. National Science Foundation, the U. S. Department of Energy, or any of the listed funding agencies.

The authors are honored to be permitted to conduct scientific research on Iolkam Du’ag (Kitt Peak), a mountain with particular significance to the Tohono O’odham Nation.



\bibliographystyle{mnras}
\bibliography{refs} % if your bibtex file is called example.bib

\bsp	% typesetting comment
\label{lastpage}
\end{document}